\documentclass[12pt,journal,draftcls,a4paper,onecolumn]{IEEEtran}
\usepackage{dsfont}
\usepackage{subfigure}
\usepackage{subeqnarray}
\usepackage{cleveref}
\usepackage{amssymb}
\usepackage{amsfonts}
\usepackage{amsmath}
\usepackage{graphicx}
\usepackage{amsmath}
\usepackage[all,poly]{xy}
\usepackage{multirow}
\usepackage{algorithm}
\usepackage{algorithmic}
\usepackage{color}
\usepackage[noadjust]{cite}

\title{Robust Linear Spectral Unmixing using Anomaly Detection}
\author{Yoann Altmann$^{(1)}$\thanks{Part of this work was supported by the Direction G\'en\'erale de l'Armement, French Ministry of Defence.}, Steve McLaughlin$^{(1)}$\thanks{Part of this work was supported by the EPSRC via grant EP/J015180/1.}, Alfred Hero$^{(2)}$\thanks{Part of this work was supported by US Army Research Office through Grant W911NF-11-1-0391.}
\thanks{Yoann Altmann and Steve
McLaughlin are with School of Engineering and Physical Sciences, Heriot-Watt University,
U.K. (email: \{y.altmann,s.mclaughlin\}@hw.ac.uk). Alfred Hero is with University of Michigan, Electrical Engineering and Computer Science, Ann Arbor, Michigan (email: hero@eecs.umich.edu).}}

\newcommand{\bLambda}{\boldsymbol{\Lambda}}
\newcommand{\bsigma}{\boldsymbol{\sigma}}
\newcommand{\bSigma}{\boldsymbol{\Sigma}}

\newcommand{\bphi}{\boldsymbol{\phi}}
\newcommand{\bbeta}{\boldsymbol{\beta}}
\newcommand{\argmax}{\operatornamewithlimits{argmax}}

\graphicspath{{figures/}}

\def\bfm{{\mathbf{m}}}

\def\bfr{{\mathbf{r}}}

\def\bfy{{\mathbf{y}}}

\def\bfC{{\mathbf{C}}}
\def\bfD{{\mathbf{D}}}
\def\bfE{{\mathbf{E}}}

\def\bfL{{\mathbf{L}}}

\def\bfR{{\mathbf{R}}}
\def\bfS{{\mathbf{S}}}

\def\bfW{{\mathbf{W}}}
\def\bfX{{\mathbf{X}}}

\def\bfZ{{\mathbf{Z}}}

\def\bbR{{\mathbb{R}}}

\newcommand{\Vpix}[1]{\mathbf{y}_{#1}}

\newcommand{\MATpix}{\mathbf{Y}}
\newcommand{\pix}[2]{y_{#1,#2}}

\newcommand{\nbpix}{N}
\newcommand{\nopix}{n}

\newcommand{\nbband}{L}

\newcommand{\nbmat}{R}
\newcommand{\nomat}{r}

\newcommand{\MATmat}{{\mathbf M}}
\newcommand{\Vmat}[1]{{\mathbf m}_{#1}}
\newcommand{\mat}[2]{m_{#1,#2}}

\newcommand{\MATabond}{{\bold A}}
\newcommand{\Vcoeff}[1]{{\bold c}_{#1}}

\newcommand{\MATcoeff}{{\bold C}}

\newcommand{\abond}[2]{{a}_{#1,#2}}
\newcommand{\Vabond}[1]{{\boldsymbol{a}}_{#1}}

\newcommand{\Vabonds}{{\boldsymbol{a}}}

\newcommand{\Vnoise}{{\mathbf e}}

\newcommand{\paramvect}{\boldsymbol{\theta}}

\newcommand{\Simplex}{\mathcal{S}}

\newcommand{\transp}{^T}

\newcommand{\etr}{\mathrm{etr}}

\newcommand{\Ndistr}[1]{\mathcal{N}\left(#1\right)}

\newcommand{\norm}[1]{\left\|#1\right\|}

\newcommand{\Vzero}{\boldsymbol{0}}
\newcommand{\Id}[1]{\textbf{I}_{#1}}
\newcommand{\Indicfun}[2]{\textbf{1}_{#1}\left(#2\right)}

\newenvironment{algogo}[1]{
\smallskip
\noindent \hrule\vspace{0.2\baselineskip} \hrule
\begin{small}
\refstepcounter{algo} \center{\bf \textsc{Algorithm \thealgo}}
\\{\center{\bf #1}}
\smallskip
\flushleft
 } {
\end{small}
\smallskip
\hrule\vspace{0.2\baselineskip} \hrule
\smallskip
}

\newcounter{algo}
\renewcommand{\thealgo}{\arabic{algo}}

\begin{document}
\maketitle

\begin{abstract}
This paper presents a Bayesian algorithm for linear spectral unmixing of hyperspectral images that accounts for anomalies present in the data. The model proposed
assumes that the pixel reflectances are linear mixtures of unknown endmembers, corrupted by an additional nonlinear term modelling anomalies and additive Gaussian
noise. A Markov random field is used for anomaly detection based on the spatial and spectral structures of the anomalies. This allows outliers to be identified in particular regions and wavelengths of the data cube.
A Bayesian algorithm
is proposed to estimate the parameters involved in the model
yielding a joint linear unmixing and anomaly detection
algorithm. Simulations conducted with synthetic and real hyperspectral images
demonstrate the accuracy of the proposed unmixing and outlier
detection strategy for the analysis of hyperspectral images.
\end{abstract}

\begin{IEEEkeywords}
Hyperspectral imagery, unsupervised spectral unmixing, Bayesian estimation, MCMC, anomaly detection.
\end{IEEEkeywords}

\section{Introduction}
Spectral unmixing (SU) of hyperspectral images (HSI) has been the subject of intensive
interest over the last two decades. It consists of distinguishing
the materials and quantifying their proportions in each pixel of an
observed image. The SU problem has been widely
studied for applications where pixel reflectances are linear
combinations of pure component spectra (called endmembers)
\cite{Heinz2001,Bioucas2012}. However,
as explained in \cite{Bioucas2012}, the linear mixing
model (LMM) can be inappropriate for some hyperspectral images, such
as those containing sand-like materials or where relief is present in the scene. Moreover, LMM-based methods can also fail when the data are corrupted by (sparse) outliers, especially when extracting the endmembers from the scene. Nonlinear
mixing models (NLMMs) have been
proposed in the hyperspectral image literature and can be divided
into two main classes \cite{Dobigeon2013spmag,Heylen2014}. The first class of NLMMs consists of physical models based on the
nature of the environment (e.g., intimate mixtures \cite{Hapke1981} and multiple scattering effects \cite{Somers2009,Nascimento2009,Halimi2010}). 
The second contains more flexible models allowing a wider range of nonlinearities to be approximated \cite{Chen2012,Altmann2014a}. 

Here, we consider a general mixing model for spectral unmixing which assumes that the observed pixels result from a convex combination
of the endmembers of the scene, corrupted by an additive term modelling deviations from the classical LMM (e.g., outliers, variability, nonlinear effects) and additive Gaussian Noise. The number of endmembers is assumed to be known whereas their spectral signatures are unknown. It is interesting to note that many nonlinear models in the literature, including polynomial models \cite{Somers2009,Nascimento2009,Halimi2010} can be expressed in a similar manner. Here, the additional terms are assumed to be \emph{a-priori} independent of the endmembers and/or their proportions (abundances), as in \cite{Dobigeon_IEEE_WHISPERS_2013,Newstadt2014ssp}. This class of models for robust linear SU allows for general deviations from the LMM to be handled in blind source separation methods, i.e., nonlinear effects, outliers or possible endmember variability \cite{Zare2014}. In \cite{Newstadt2014ssp}, spatial and spectral sparsity structures were considered for the additional term since deviations from the LMM can occur in specific regions or spectral bands of the HSI. This is typically the case when outliers are present, but also when nonlinear effects (relief) occurs and when the reflectance of materials present has significant variations in particular spectral ranges (e.g. due to natural variability of vegetation). In this paper, we extend \cite{Newstadt2014ssp} by introducing a probabilistic 3D Ising model for spatial and spectral influence of outliers thus allowing for more flexible group-sparsity structures for the support sets of anomalies, whereas \cite{Newstadt2014ssp} assumed the support sets of outliers to have a fixed structure. Moreover, the algorithm presented in this paper allows the estimation of the Ising model parameters directly from the data.

Adopting a Bayesian framework, we assign prior distributions to the unknown model parameters to include available information (such as parameter constraints) within the estimation procedure. In particular, an Ising Markov random field is introduced to model spatial and spectral correlations for the anomalies. The joint posterior distribution of the unknown parameter vector is then derived. Since classical Bayesian estimators cannot be easily computed from this joint posterior,
a Markov chain Monte Carlo (MCMC) method is used to generate samples according to this posterior. More precisely, we construct an efficient stochastic gradient MCMC (SGMCMC) algorithm \cite{Pereyra2014ssp} that simultaneously
estimates the endmember and abundance matrices along with the Ising hyperparameters.

The main contributions of this work are threefold:
\begin{enumerate}
\item We develop a new hierarchical outlier model taking into account spatial and spectral correlations through Markovian dependencies, this contrasts with the model proposed in \cite{Newstadt2014ssp} which considered a fixed outlier structure. This flexible model is embedded within the mixing model for robust unsupervised linear spectral unmixing \emph{via} anomaly detection.
\item An adaptive MCMC algorithm is
proposed to compute the Bayesian estimates of interest and
perform Bayesian inference. This algorithm is equipped with a
stochastic optimisation adaptation mechanism that automatically
adjusts the parameters of the Markov random field by
maximum marginal likelihood estimation, thus removing the need to set the regularisation parameters by cross-validation.
\item We show the benefits of the proposed flexible model for linear spectral unmixing of synthetic and real hyperspectral images. Specifically, we demonstrate the ability of the proposed algorithm to detect structured anomalies thus enhancing endmember and abundance estimation.
\end{enumerate}

The remaining sections of the paper are organized as follows. Section
\ref{sec:Problem} introduces the mixing model for robust linear SU of HSIs, followed by Section \ref{sec:bayesian} which summarizes the likelihood and the priors assigned to the unknown parameters of the model. The resulting joint posterior distribution and the Gibbs sampler used to
sample from it are summarized in Section
\ref{sec:Gibbs}. A generalization of the proposed Bayesian model for robust Bayesian subspace identification is proposed in Section \ref{sec:RBPCA}. Some simulation results conducted on
synthetic data are shown and discussed in Section \ref{sec:simu}. Conclusions and future work are reported in Section
\ref{sec:conclusion}.

\vspace{-0.3cm}
\section{Problem formulation}
\label{sec:Problem} We consider a set of $N$ observed pixels/spectra
$\Vpix{n} =[\pix{1,n},\ldots,\pix{\nbband}{n}]\transp, n \in \left
\lbrace 1,\ldots,N \right \rbrace$ where $\nbband$ is the number of
spectral bands. Each of these spectra is assumed to result from a linear
combination of $R$ unknown endmembers $\Vmat{r}$, corrupted by possible additive outliers and Gaussian noise. The observation model can be expressed as
\begin{eqnarray}
\label{eq:NLM0}
 \Vpix{n} &= & \sum_{r=1}^R \Vmat{r} \abond{r}{n} + \bfr_n + \Vnoise_{n} \nonumber\\
 & = & \MATmat \Vabond{n} + \bfr_n + \Vnoise_{n}, \quad n=1,\ldots,N
\end{eqnarray}
where $\Vmat{\nomat} =
[\mat{\nomat}{1},\ldots,\mat{\nomat}{\nbband}]\transp$ is the
spectrum of the $\nomat$th material present in the scene and
$\abond{r}{n}$ is its corresponding proportion (abundance) in the $n$th pixel. In \eqref{eq:NLM0}, 
$\Vnoise_n$ is an additive independently but non identically distributed
zero-mean Gaussian noise sequence with diagonal covariance matrix
$\bSigma_0=\textrm{diag}(\bsigma^2)$, denoted as $\Vnoise_n \sim
\Ndistr{\Vnoise_n;\Vzero_{\nbband},\bSigma_0}$, where
$\bsigma^2=[\sigma_1^2,\ldots,\sigma_L^2]\transp$ is the vector of
the $L$  noise variances and $\textrm{diag}(\bsigma^2)$ is an $L \times L$
diagonal matrix containing the elements of the vector $\bsigma^2$. Moreover, $\bfr_n$ denotes the outlier vector of the $n$th pixel. Note
that the usual matrix and vector notations $\MATmat =
[\Vmat{1},\ldots,\Vmat{\nbmat}]$ and
$\Vabond{n}=[\abond{1}{n},\ldots, \abond{\nbmat}{n}]\transp$ have
been used in the second row of \eqref{eq:NLM0}.

As a consequence of physical constraints, the abundance vectors $\Vabond{n}$
satisfy the following positivity and sum-to-one constraints

\begin{eqnarray}
\label{eq:abundancesconst}
\sum_{\nomat=1}^{\nbmat}{\abond{r}{n}}=1,~~ \abond{r}{n} >
0, \forall \nomat \in \left\lbrace 1,\ldots,\nbmat \right\rbrace.
\end{eqnarray}

The problem investigated in this paper is to estimate the endmember matrix $\MATmat$, the abundance matrix $\MATabond=[\Vabond{1},\ldots,\Vabond{N}]$, the noise variances in $\bsigma^2$ and the outlier matrix $\bfR=[\bfr_1\ldots,\bfr_N]$ from the observation matrix $\MATpix=[\Vpix{1},\ldots,\Vpix{N}]$. 
To solve this problem, we propose a hierarchical Bayesian model and a sampling method to estimate the unknown parameters. 

\vspace{-0.3cm}
\section{Robust Bayesian Linear Unmixing (RBLU)}
\label{sec:bayesian}
\subsection{Likelihood}
Eq. \eqref{eq:NLM0} implies that $\Vpix{n}|\MATmat,\Vabond{n},
\bfr_{n}, \bsigma^2 \sim
\Ndistr{\Vpix{n};\MATmat\Vabond{n}+\bfr_n,\bSigma_0}$. Assuming independence
between noise sequences of the $N$ observed pixels, the likelihood of the observation matrix 
$\MATpix$ can be expressed as\\
$ f(\MATpix|\MATmat,\MATabond, \bfR, \bsigma^2) \propto$
\begin{eqnarray}
\label{eq:likelihood}
  |\bSigma_0|^{-N/2}\etr\left[-\dfrac{(\MATpix-\MATmat\MATabond -\bfR)\transp\bSigma_0^{-1}(\MATpix-\MATmat\MATabond -\bfR)}{2}\right]
\end{eqnarray}
where $\propto$ means ``proportional to'' and
$\etr(\cdot)$ denotes the exponential trace.
\subsection{Parameter priors}

\subsubsection{Prior for the abundance matrix $\MATabond$}
Each abundance vector can be written as
$\Vabond{n}=[\Vcoeff{n}\transp,\abond{R}{n}]\transp$
with $\Vcoeff{n}=[\abond{1}{n},\ldots,
\abond{\nbmat-1}{n}]\transp$ and
$\abond{R}{n}=1-\sum_{\nomat=1}^{\nbmat-1}{\abond{r}{n}}$. The LMM
constraints \eqref{eq:abundancesconst} impose that $\Vcoeff{n}$
belongs to the simplex $\Simplex = \left\lbrace \Vcoeff{} \left|
c_{\nomat}\geq 0,\forall r\in 1,\ldots,R-1,
\sum_{\nomat=1}^{\nbmat-1}{c_{\nomat}} \leq 1 \right\rbrace
\right.$. To reflect the lack of prior knowledge about
the abundances, a uniform prior is assigned for each vector $\Vcoeff{n}, n \in \left \lbrace
1,\ldots,N\right \rbrace$, i.e., 
$f(\Vcoeff{n}) \propto
\Indicfun{\Simplex}{\Vcoeff{n}}$,
where $\Indicfun{\Simplex}{\cdot}$ is the indicator function defined on the simplex $\Simplex$. When prior knowledge about the abundances are available, the uniform prior can be replaced by more informative priors such as (mixtures of) Dirichlet distributions \cite{Nascimento2012} or Gaussian mixtures using logistic coefficients \cite{Eches2011}. Assuming prior independence
between the $N$ abundance vectors $\left \lbrace \Vabond{n} \right
\rbrace_{n=1,\ldots,N}$ leads to the following joint prior
distribution
\begin{eqnarray}
\label{eq:prior_C}
f(\MATcoeff)  =
\prod_{n=1}^{N}f(\Vcoeff{n}),
\end{eqnarray}
where $\MATcoeff=[\Vcoeff{1},\ldots,\Vcoeff{N}]$ is an $(R-1)
\times N$ matrix.

\subsubsection{Prior for the endmember matrix $\MATmat$}
To reflect the lack of prior knowledge about the endmembers, 
we use the following multivariate truncated Gaussian prior 
\begin{eqnarray}
\label{eq:prior_M_star}
f(\MATmat|\xi) \propto \prod_{r=1}^R \mathcal{N}_{\left(\bbR^+\right)^L}(\Vmat{r};\Vzero,\xi\Id{L})
\end{eqnarray}
where $\xi$ is fixed to a large value, to ensure endmember positivity while using a weakly informative prior. Note that \eqref{eq:prior_M_star} is considered in order to handle the case where the data are not normalized. If the data are actual reflectance values, a prior can be introduced ensuring that the endmember spectra belong to $(0,1)$, such as a uniform, beta \cite{Du2014} or Gaussian distribution. Note that the prior can also include prior information from an endmember extraction algorithm, as in \cite{Dobigeon2009,Altmann2014b}.

\subsubsection{Prior for the noise variances}
A Jeffreys' prior is chosen for the noise variance in each spectral
band $\sigma_{\ell}^2$, i.e., $f(\sigma_{\ell}^2) \propto \sigma_{\ell}^{-2} \Indicfun{\bbR^+}{\sigma_{\ell}^2}$ where $\Indicfun{\bbR^+}{\cdot}$ denotes the indicator function defined on $\bbR^+$, which reflects the absence of knowledge about these parameters. Again, these non-informative priors can be easily replaced by conjugate inverse-Gamma priors to include prior knowledge available about the noise levels. Assuming prior independence
between the noise variances, we obtain
\begin{eqnarray}
f(\bsigma^2) = \prod_{\ell=1}^L f(\sigma_{\ell}^2).
\end{eqnarray}

\subsubsection{Priors of the outliers}
As in \cite{Newstadt2014ssp,Altmann2014a}, the outliers are assumed to be sparse, i.e., at most of the pixels and spectral bands, the outliers are expected to be exactly equal to zero. To model the outlier sparsity, we factorize the outlier matrix as 
\begin{eqnarray}
\label{eq:outliers_decomp}
\bfR = \bfZ \odot \bfX,
\end{eqnarray}
where $\bfZ \in \left\lbrace 0,1 \right\rbrace^{\nbband \times N}$ is a label matrix, $\bfX \in \bbR^{\nbband \times N}$ and $\odot$ denotes the Hadamard (termwise) product. This decomposition allows one to decouple the location of the sparse components from their values. More precisely, 
$z_{\ell,n}=\left[\bfZ\right]_{\ell,n}=1$ if an outlier is present in the $\ell$th spectral band of the $n$th observed pixel with value equal to $r_{\ell,n}=x_{\ell,n}$. A conjugate Gaussian prior is used for $\bfX$, i.e., 
\begin{eqnarray}
\label{eq:prior_X}
f(\bfX|s^2) = \prod_{\ell,n} \Ndistr{x_{\ell,n};0,s^2},
\end{eqnarray}
where $s^2$ controls the prior energy of the outliers. Note that \eqref{eq:prior_X} allows the outliers to be negative. Other conjugate priors, such as exponential or truncated Gaussian priors, could be used instead of \eqref{eq:prior_X}, e.g., to enforce outlier positivity. The next paragraph presents the prior considered for the label matrix $\bfZ$.

\subsubsection{Label matrix}
For many applications, the locations of outliers are likely to be spectrally (e.g., water absorption bands) and/or spatially (e.g. weakly represented components, shadowing effects) correlated. An effective way to take correlated outliers/nonlinear effects into account is to consider Markov random fields (MRF) to build a prior for the label matrix $\bfZ$ \cite{Altmann2014a}. MRFs assume that the distribution of a label $z_{\ell,n}$ conditionally to the other labels of
the image equals the distribution of this label vector conditionally to its neighbors, i.e., $\textrm{P}(z_{\ell,n}|\bfZ_{\backslash z_{\ell,n}})= \textrm{P}(z_{\ell,n}|\bfZ_{\mathcal{V}_{\ell,n}})$, where $\mathcal{V}_{\ell,n}$ is the index set of the neighbors of $z_{\ell,n}$, 
$\bfZ_{\backslash z_{\ell,n}}$
denotes the matrix $\bfZ$
whose element $z_{\ell,n}$ has been removed and
$\bfZ_{\mathcal{V}_{\ell,n}}$ 
is the subset of $\bfZ$ composed of the elements whose indexes belong to $\mathcal{V}_{\ell,n}$. 
In this study, we consider that the spatial and spectral correlations can be different and thus consider two different neighborhoods. We decompose the neighborhood $\mathcal{V}_{\ell,n}$ as 
$\mathcal{V}_{\ell,n} = \mathcal{V}_{\ell,n}^L \cup \mathcal{V}_{\ell,n}^N$ where $\mathcal{V}_{\ell,n}^N$ (resp. $\mathcal{V}_{\ell,n}^L$) denotes the spatial (resp. spectral) neighborhood of $z_{\ell,n}$.
In this paper, we consider an Ising model that can be expressed as 
\begin{eqnarray}
\label{eq:Potts}
\textrm{P}(\bfZ | \bbeta') & = & \dfrac{1}{B(\bbeta')} \exp \left[\bbeta\transp \bphi(\bfZ) + \phi_{0}\left(\bfZ,\beta_0\right)\right]
\end{eqnarray}
where $\bbeta=[\beta_{N},\beta_{L}]\transp$, $\bbeta'=[\bbeta\transp,\beta_0]\transp$ and
\begin{eqnarray}
\left\{
    \begin{array}{lll}
        \phi_{L}\left(\bfZ\right) & = & \sum_{n,\ell} \sum_{z_{\ell',n} \in \mathcal{V}_{\ell,n}^L} \delta(z_{\ell,n} - z_{\ell',n}),\nonumber\\
\phi_{N}\left(\bfZ\right) & = & \sum_{n,\ell} \sum_{z_{\ell,n'} \in \mathcal{V}_{\ell,n}^N} \delta(z_{\ell,n} - z_{\ell,n'}),\nonumber\\
\bphi(\bfZ) & = & [\phi_{L}\left(\bfZ\right),\phi_{N}\left(\bfZ\right)]\transp,\nonumber\\
\phi_{0}\left(\bfZ,\beta_0\right) & = & \beta_0\sum_{n,\ell}  (1-z_{\ell,n}) +(1-\beta_0)\sum_{n,\ell}  z_{\ell,n},\nonumber

    \end{array}
\right.
\end{eqnarray}
and $\delta(\cdot)$ denotes the Kronecker delta function. Moreover, $\beta_{N}>0$ and $\beta_{L}>0$ are hyperparameters that control the spatial and spectral granularity of the MRF and $0\leq \beta_0 \leq 1$ is an additional parameter that models the probability of having outliers in the image. Specifically, the higher the value of $\beta_0$, the lower the probability of outliers in the data. The estimation of the proposed Ising model hyperparameters will be discussed in the next section. Different spectral and spatial neighbourhoods can be used in \eqref{eq:Potts}. In this paper, we consider a 4-neighbour structure to account for the spatial correlation and a 2-neighbour structure for the spectral dimension.  

\subsection{Outlier variance $s^2$}
The following conjugate inverse-Gamma prior is assigned to $s^2$
\begin{eqnarray}
\label{eq:prior_s2}
s^2 \sim \mathcal{I}\mathcal{G} (\gamma,\nu),
\end{eqnarray}
where $(\gamma,\nu)$ are fixed to $(\gamma,\nu)=(10^{-3},10^{-3})$ to ensure a weakly informative prior. This choice of hyperparameters reflects the lack of prior information about the outliers variance. However, \eqref{eq:prior_s2} could be replaced by more informative priors by suitably adapting $(\gamma,\nu)$. 

\subsection{Joint posterior distribution}
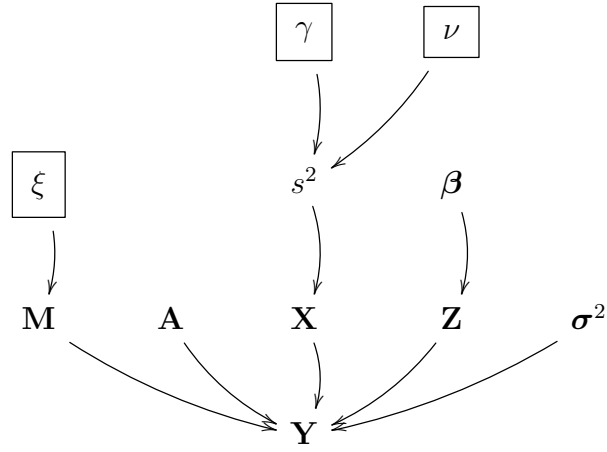
\begin{figure}[!ht]
\centerline{ \xymatrix{
 & & *+<0.05in>+[F-]+{\gamma} \ar@/^/[d] & *+<0.05in>+[F-]+{\nu} \ar@/^/[ld] & \\
 *+<0.05in>+[F-]+{\xi} \ar@/^/[d] &  & s^2 \ar@/^/[d]& \bbeta \ar@/^/[d] & \\
  \MATmat \ar@/_/[rrd]    & \MATabond \ar@/_/[rd] &   \bfX \ar@/^/[d]& \bfZ \ar@/^/[ld] & \bsigma^2 \ar@/^/[lld]  \\
   & & \MATpix &   & }
} \caption{Directed acyclic graph representing the proposed hierarchical Bayesian model (fixed quantities appear in boxes).} \label{fig:DAG}
\end{figure}

Assuming the parameters $\MATmat, \MATabond, \bfZ, \bfX$ and $\bsigma^2$ are a priori independent, the joint posterior of the parameter vector $\paramvect=\left\lbrace  \MATmat, \MATabond, \bfX, \bfZ,\bsigma^2 \right \rbrace$ and the parameter $s^2$ can be expressed as 
\begin{eqnarray}
\label{eq:posterior}
f(\paramvect, s^2|\MATpix,\bphi,\bbeta') \propto f(\MATpix|\paramvect)f(\paramvect|s^2,\xi,\bbeta)f(s^2|\gamma,\nu)
\end{eqnarray}
where 
\begin{eqnarray}
\label{eq:joint_post}
f(\paramvect|s^2,\xi,\bbeta') = f(\MATmat|\xi)f(\MATabond)f(\bsigma^2)f(\bfX|s^2)\textrm{P}(\bfZ|\bbeta')
\end{eqnarray}
and $\bphi=[\xi,\gamma,\nu]\transp$ a vector of model hyperparameters. The MRF parameter vector $\bbeta'$ will be determined by maximum marginal likelihood estimation during the inference
procedure.
The directed acyclic graph (DAG) summarizing the structure of proposed Bayesian model is depicted in Fig. \ref{fig:DAG}.

Next we describe an MCMC method for sampling from this posterior distribution to estimate the unknown model parameters.

\vspace{-0.3cm}
\section{Bayesian Inference}
\label{sec:Gibbs}
The Bayesian model defined in Section \ref{sec:bayesian} specifies the joint
posterior density for the unknown parameters $\paramvect, s^2$
given the observations $\MATpix$ and the hyperparameters $\xi,\gamma,\nu$ and $\bbeta'$. 
This posterior distribution models the complete knowledge about the unknown parameters given the observed data and the
prior information available. We propose the following Bayesian estimators for hyperspectral unmixing and nonlinearity detection: the marginal posterior mean or
minimum mean square error estimator of the abundance and endmember matrices
\begin{eqnarray}
\label{abund_endm_MMSE}
\left(\widehat{\MATabond}_{\textrm{MMSE}},\widehat{\MATmat}_{\textrm{MMSE}}\right) =  \textrm{E}\left[\MATabond,\MATmat \left| \MATpix,\bphi,\hat{\bbeta}'\right.\right],
\end{eqnarray}
where the expectation is taken with respect to the marginal posterior density $f(\MATabond,\MATmat | \MATpix,\bphi,\hat{\bbeta}')$ (by marginalizing out $\bfZ,\bfX,\bsigma^2$ and $s^2$, this density takes into account their uncertainty), the marginal maximum \emph{a posteriori} (MMAP) estimator for the outlier support $\bfZ$
\begin{eqnarray}
\label{zEstimator}
z^{MMAP}_{n,\ell} = \argmax_{z_{n,\ell} \in \{0,1\}} f (z_{n,\ell} |\MATpix,\bphi,\hat{\bbeta}'),
\end{eqnarray}
and, conditionally on the estimated outliers location, the minimum mean square error estimator of the outlier values
\begin{eqnarray}
\label{rEstimator}
r^{MMSE}_{n,\ell} = \textrm{E}\left[ x_{n,\ell} \left| z_{n,\ell} = z^{MMAP}_{n,\ell},  \MATpix, \bphi,\hat{\bbeta}' \right.\right],
\end{eqnarray}
where 
\begin{eqnarray}
f (z_{n,\ell}|\MATpix,\bphi,\hat{\bbeta}') = \int f(\paramvect,s^2| \MATpix,\bphi,\hat{\bbeta}') \textrm{d}\paramvect_{\backslash z_{n,\ell}} \textrm{d}s^2,
\end{eqnarray}
where $\textrm{E}\left[\cdot\right]$ denotes the expectation with respect to the conditional marginal density
\begin{eqnarray}
f (x_{n,\ell} | z_{n,\ell},  \MATpix, \bphi,\hat{\bbeta}')\\ = \frac{\int f(\paramvect,s^2| \MATpix, \MATpix, \bphi,\hat{\bbeta}') \textrm{d}\paramvect_{\backslash z_{n,\ell}}\textrm{d}s^2}{f (z_{n,\ell} |\MATpix, \bphi,\hat{\bbeta}')}.
\end{eqnarray}
Note that the outlier estimator \eqref{rEstimator} is sparse by construction (\emph{i.e.}, $\textrm{E}\left[ x_{n,\ell} \left| z_{n,\ell} = 0,   \MATpix, \bphi,\hat{\bbeta}'\right.\right] = 0$).

Computing \eqref{abund_endm_MMSE}, \eqref{zEstimator} and \eqref{rEstimator} is challenging because it requires access to the joint marginal density of $(\MATmat,\MATabond)$, the univariate marginal densities of $z_{n,\ell}$ and the joint marginal densities of $(x_{n,\ell}, z_{n,\ell})$, which in turn require computing the posterior \eqref{eq:posterior} and performing an integration over a very high-dimensional space. Fortunately, these can be efficiently approximated with arbitrarily good accuracy by Monte Carlo integration. More precisely, it is possible to compute \eqref{abund_endm_MMSE}, \eqref{zEstimator} and \eqref{rEstimator}  by first using an MCMC computational method to generate samples asymptotically distributed according to \eqref{eq:posterior}, and subsequently using these samples to approximate the required marginal probabilities and expectations. Note that in \eqref{abund_endm_MMSE}, \eqref{zEstimator} and \eqref{rEstimator}, we have set $\bbeta' = \hat{\bbeta}'$, which denotes the maximum marginal likelihood estimator of the Ising regularisation hyperparameter vector $\bbeta'$ given the observed data $\MATpix$, i.e.,
\begin{eqnarray}\label{betaML}
\hat{\bbeta}'=\underset{\bbeta' \in \mathcal{B}}{\textrm{argmax}} f\left(\MATpix | \bphi,\bbeta'\right).
\end{eqnarray}
This is an empirical Bayes approach for specifying $\bbeta'$ where hyperparameters with unknown values are replaced by point estimates computed from observed data (as opposed to being fixed a priori or integrated out of the model by marginalisation). As explained in \cite{Pereyra2014ssp}, this strategy has several important advantages for MRF hyperparameters such as $\bbeta'$ having intractable conditional distributions. In particular, it allows automatic adjustment of the value of $\bbeta'$ to each image, producing significantly better estimation results than using a single fixed value of $\bbeta'$ for all images. Furthermore it has significantly lower computational cost compared to that of competing approaches, such as including $\bbeta'$ in the model and subsequently marginalizing it during the inference procedure \cite{Pereyra2013ip}.

\subsection{Bayesian estimation algorithm}
Exact computation of the MMSE and MMAP estimators \eqref{abund_endm_MMSE}, \eqref{zEstimator} and \eqref{rEstimator} is very challenging because it involves calculating expectations with respect to posterior marginal densities, which in turn require evaluating the full posterior \eqref{eq:posterior} and integrating it over a very high-dimensional space. Exact computation of $\hat{\bbeta}'$ is also difficult because it involves solving an intractable optimisation problem (it is not possible to evaluate the exact marginal likelihood $f(\MATpix | \alpha_3)$ or its gradient $\nabla f(\MATpix | \alpha_3)$). Here we follow the approach proposed in \cite{Pereyra2014ssp} and design a stochastic optimisation and simulation algorithm to compute \eqref{abund_endm_MMSE}, \eqref{zEstimator} and \eqref{rEstimator} simultaneously. We construct an SGMCMC algorithm that simultaneously estimates $\hat{\bbeta}'$ and generates a chain of $N_{\textrm{MC}}$ samples $\{\MATmat^{(t)},\MATabond^{(t)}\}_{t=1}^{N_{\textrm{MC}}}$ asymptotically distributed according to the marginal density $f(\MATmat,\MATabond | \MATpix,\hat{\bbeta}')$ (this algorithm is summarised in Algorithm \ref{algo:algo1} below). Once the samples have been generated, the estimators \eqref{abund_endm_MMSE}, \eqref{zEstimator} and \eqref{rEstimator} are approximated by Monte Carlo integration \cite[Chap. 10]{Robert2004}, i.e.,
\begin{eqnarray}
\label{eq:abund_endm_MC}
\hat{\MATmat}_{MMSE} & = & \dfrac{1}{N_{\textrm{MC}}-N_{\textrm{bi}}}\sum_{t=N_{\textrm{bi}}+1}^{N_{\textrm{MC}}}
	\MATmat^{(t)},\nonumber\\
\hat{\MATabond}_{MMSE} & = & \dfrac{1}{N_{\textrm{MC}}-N_{\textrm{bi}}}\sum_{t=N_{\textrm{bi}}+1}^{N_{\textrm{MC}}}
	\MATabond^{(t)},\nonumber\\
z^{MMAP}_{n,\ell} & = & \left\lbrace
\begin{array}{ll}
 0 & \mbox{if} ~~\textrm{card}(\mathcal{Z}_{n,\ell})\leq \left(N_{\textrm{MC}}-N_{\textrm{bi}}\right)/2\\
 1 & \mbox{else}, \\
\end{array}
\right.\nonumber
\end{eqnarray}
$r^{MMSE}_{n,\ell} =$ 
\begin{eqnarray}
\left\lbrace
\begin{array}{ll}
 \dfrac{1}{\textrm{card}(\mathcal{Z}_{n,\ell})}\sum_{z_{n,\ell}^{(t)}\in \mathcal{Z}_{n,\ell}}
	r^{(t)}_{n,\ell} & \mbox{if} ~~ z^{MMAP}_{n,\ell}=1\\
 0 & \mbox{else}, \nonumber\\
\end{array}
\right.
\end{eqnarray}
with $\mathcal{Z}_{n,\ell}=\left\lbrace z_{n,\ell}^{(t)}| t \in \{N_{\textrm{bi}}+1,\ldots,N_{\textrm{MC}} \}, z_{n,\ell}^{(t)}=1\right\rbrace$ and where the samples from the first $N_{\textrm{bi}}$ iterations (corresponding to the transient regime or burn-in period) are discarded. The main steps of this algorithm are detailed in below.

\subsubsection{Sampling the labels}
It can be seen from \eqref{eq:posterior} that 
\begin{eqnarray}
f(z_{n,\ell}=i|\MATpix,\paramvect_{\backslash z_{n,\ell}},\bbeta',s^2) \propto \tilde{\pi}_{n,\ell}^{(i)},\quad \forall (n,\ell),
\end{eqnarray}
where $i \in \{0,1\}$ and 
\begin{eqnarray}
\log\left(\tilde{\pi}_{n,\ell}^{(i)}\right) & = &-\dfrac{\left(\pix{\ell}{n}-\Vmat{\ell,:}\Vabond{n} - i x_{\ell,n}\right)^2}{2\sigma_{\ell}^2} \nonumber\\
& - & \bbeta\transp \bphi(\bfZ) - \phi_{0}\left(\bfZ,\beta_0\right).
\end{eqnarray}
Consequently, the label $z_{n,\ell}$  can be drawn from its conditional distribution by drawing randomly from $\{0,1\}$ with probabilities given by 
\begin{eqnarray}
\label{eq:post_Z}
f(z_{n,\ell}=i|\MATpix,\paramvect_{\backslash z_{n,\ell}},\bbeta',s^2) = \dfrac{\tilde{\pi}_{n,\ell}^{(i)}}{\tilde{\pi}_{n,\ell}^{(0)}+\tilde{\pi}_{n,\ell}^{(1)}}.
\end{eqnarray}

\subsubsection{Sampling the endmembers}
It can be easily shown that 
\begin{eqnarray}
\label{eq:joint_post_M}
f(\MATmat|\MATpix,\paramvect_{\backslash \MATmat},s^2,\bphi,\bbeta') = \prod_{\ell=1}^L f(\Vmat{\ell,:}|\MATpix,\paramvect_{\backslash \Vmat{\ell,:}},s^2,\xi),
\end{eqnarray}
i.e., the rows of $\MATmat$, denoted as $\{\Vmat{\ell,:}\}$ are \emph{a posteriori} independent (conditioned on the other parameters). Moreover,
\begin{eqnarray}
\label{eq:post_M}
\Vmat{\ell,:}\left|\MATpix,\paramvect_{\backslash \Vmat{\ell,:}},s^2,\xi \sim \mathcal{N}_{\left(\bbR^+\right)^R}\left(\Vmat{\ell,:}; \tilde{\bfm}_{\ell,:},\bfS_{\ell}^{(\MATmat)}\right)\right..
\end{eqnarray}
where
\begin{eqnarray}
\tilde{\bfm}_{\ell,:} &= &\sigma_{\ell}^{-2} \tilde{\bfy}_{\ell,:}\MATabond\transp \bfS_{\ell}^{(\MATmat)}\nonumber\\
\bfS_{\ell}^{(\MATmat)}& = &\left(\sigma_{\ell}^{-2}\MATabond\MATabond\transp + \xi^{-2}\Id{L}\right)^{-1}
\end{eqnarray}
and $\tilde{\bfy}_{\ell,:}$ is the $\ell$th row of the $L \times N$ matrix $\widetilde{\MATpix}=\MATpix-\bfR$.
Sampling from \eqref{eq:post_M} can be achieved efficiently by using the Hamiltonian method recently proposed in
\cite{Pakman2012} or by successive sampling from $R$ truncated Gaussian distributions (via Gibbs sampling).

\subsubsection{Sampling the abundances}
In a similar fashion to obtaining \eqref{eq:joint_post_M}, it can be easily shown that 
\begin{eqnarray}
\label{eq:post_A}
f(\bfC|\MATpix,\paramvect_{\backslash \bfC},s^2,\bphi,\bbeta') = \prod_{n=1}^N f(\Vcoeff{n}|\Vpix{n},\paramvect_{\backslash \Vcoeff{n}},s^2),
\end{eqnarray}
i.e., the columns of $\MATabond$ are \emph{a posteriori} independent (conditioned on the other parameters). 
Moreover, the conditional distribution of $\Vcoeff{n}|\Vpix{n},\paramvect_{\backslash \Vcoeff{n}},s^2$ is a multivariate 
Gaussian distribution restricted to the simplex $\Simplex$, which
can be sampled efficiently using the method proposed in\cite{Pakman2012}.

\subsubsection{Sampling the latent variable matrix $\bfX$}
In a similar manner to the abundance matrix in \eqref{eq:post_A}, the elements of $\bfX$ are \emph{a posteriori} independent (conditioned on the other parameters) and can be sampled independently. Since the prior \eqref{eq:prior_X} is conjugate to the Gaussian distribution, the full conditional distribution of $x_{\ell,n}$ is given by
\begin{eqnarray}
\label{eq:post_X}
x_{\ell,n}\left|\pix{\ell}{n},\paramvect_{\backslash x_{\ell,n}},s^2,\bbeta'\sim \Ndistr{x_{\ell,n};\tilde{x}_{\ell,n},\tilde{\sigma}_{\ell,n}^2}\right.
\end{eqnarray}
where 
\begin{eqnarray}
\tilde{x}_{\ell,n} &= &z_{\ell,n}(\pix{\ell}{n}-\Vmat{\ell,:}\Vabond{n})\dfrac{\tilde{\sigma}_{\ell,:}^2}{\sigma_{\ell}^2}\nonumber\\
\tilde{\sigma}_{\ell,n}^2& = &\dfrac{\sigma_{\ell}^2s^2}{\sigma_{\ell}^2+ z_{\ell,n}s^2}.
\end{eqnarray}

\subsubsection{Sampling the noise variances}
Sampling the noise variances can be easily achieved by sampling from the following $L$ independent inverse-Gamma distributions
\begin{eqnarray}
\label{eq:post_sigma2}
\sigma_{\ell}^2\left|\MATpix,\paramvect_{\sigma_{\ell}^2},s^2\sim \mathcal{IG}\left(\sigma_{\ell}^2;\frac{N}{2},\frac{\norm{\Vpix{\ell,:}-\Vmat{\ell,:}\MATabond -\bfr_{\ell,:}}^2}{2} \right) \right.
\end{eqnarray}

\subsubsection{Sampling the outlier variance $s^2$} 
Finally, in a similar fashion to the noise variances, it can be shown that 
\begin{eqnarray}
\label{eq:post_s2}
s^2\left|\MATpix,\paramvect,\bphi \sim \mathcal{IG}\left(\dfrac{NL}{2}+ \gamma, \nu+ \sum_{n,\ell}\frac{x_{\ell,n}^2}{2} \right)\right.
\end{eqnarray}

\subsubsection{Updating the Ising regularisation model parameter vector $\bbeta'$}
If the marginal likelihood $f(\MATpix | \bphi,\bbeta')$ was tractable, we could update $\bbeta'$ from one MCMC iteration to the next by using a classic gradient descent step
\begin{eqnarray*}
\bbeta'^{(t)} = \bbeta'^{(t-1)} + \delta_t \nabla \log f(\MATpix | \bphi,\bbeta'^{(t-1)}),
\end{eqnarray*}
with $\delta_t = t^{-3/4}$, such that $\bbeta'^{(t)}$ converges to $\hat{\bbeta}'$ as $t \rightarrow \infty$. However, this gradient has two levels of intractability, one due to the marginalisation of $(\paramvect,s^2)$ and another one due to the intractable normalising constant of the Ising model. We address this difficulty by following the approach proposed in \cite{Pereyra2014ssp}; that is, by replacing $\nabla \log f(\MATpix | \bphi,\bbeta'^{(t)})$ with an estimator computed with the samples generated by the MCMC algorithm at iteration $t$, and a set of two auxiliary variables  $(\bfZ') \sim \mathcal{K}(\bfZ|\bfZ^{(t)},\bbeta'^{(t-1)})$ generated with an MCMC kernel $\mathcal{K}$ with target density \eqref{eq:Potts} (in our experiments we used a Gibbs sampler implemented using a colouring scheme such that half of the elements of $\bfZ'$ are generated in parallel). The updated value $\bbeta'^{(t)}$ is then projected onto the domain $\mathcal{B}_t=[0,B_t]\times[0,B_t]\times[0,1]$ to guarantee the constraints of  $\beta_N, \beta_L$ and $\beta_0$ and the stability of the stochastic optimisation algorithm, where $B_t$ is an arbitrarily large upper bound on $\beta_N$ and $\beta_L$ that can be
increased at every iteration. In our experiments we have used $B_t=10$.

It is worth mentioning that if it was possible to simulate the auxiliary variables $\bfZ'$ exactly from \eqref{eq:Potts} then the estimator of $\nabla \log f(\MATpix | \bphi,\bbeta'^{(t-1)})$ used in Algorithm \ref{algo:algo1} would be unbiased and as a result $\bbeta'^{(t)}$ would converge exactly to $\hat{\bbeta}'$. However, exact simulation from \eqref{eq:Potts} is not computationally feasible and therefore we resort to the MCMC kernel $\mathcal{K}$ and obtain a biased estimator of $\nabla \log f(\MATpix | \bphi,\bbeta'^{(t-1)})$ that drives $\bbeta'^{(t)}$ to a neighbourhood of $\hat{\bbeta}'$ \cite{Pereyra2014ssp}. We found that computing this is significantly less expensive than alternative approaches, e.g., using an approximate Bayesian computation algorithm \cite{Pereyra2013ip}, and that it leads to very accurate unmixing results.
\begin{algogo}{RBLU algorithm}
     \label{algo:algo1}
     \begin{algorithmic}[1]
        \STATE \underline{Fixed input parameters:} Number of endmembers $R$, number of burn-in iterations $N_{\textrm{bi}}$, total number of iterations $N_{\textrm{MC}}$
				\STATE \underline{Initialization ($t=0$)}
        \begin{itemize}
        \item Set $\MATabond^{(0)},\MATmat^{(0)},\bfX^{(0)},\bfZ^{(0)},\bsigma^{2(0)},s^{2(0)},\bbeta^{(0)}$
        \end{itemize}
        \STATE \underline{Iterations ($1 \leq t \leq N_{\textrm{MC}}$)}
				\STATE Sample $\bfZ^{(t)}$ from \eqref{eq:post_Z}
        \STATE Sample $\MATmat^{(t)}$ from \eqref{eq:post_M}
        \STATE Sample $\MATabond^{(t)}$ from \eqref{eq:post_A}
        \STATE Sample $\bfX^{(t)}$ from \eqref{eq:post_X}
				\STATE Sample $\bsigma^{2(t)}$ from \eqref{eq:post_sigma2}
				\STATE Sample $s^{2(t)}$ from \eqref{eq:post_s2}
				\IF{$t<N_{\textrm{bi}}$} 
				\STATE Sample $\bfZ' \sim \mathcal{K}(\bfZ|\bfZ^{(t)},\bbeta'^{(t-1)})$
				\STATE Set $\beta_N^{(t)}=\mathcal{P}_{\left[0,B_t\right]}\left(\beta_N^{(t-1)} + \Delta \beta_N\right)$ with 
				\STATE $\Delta \beta_N = \delta_t\left[\frac{\textrm{d}}{\textrm{d}\beta_N} \log f(\bfZ|\bbeta') - \frac{\textrm{d}}{\textrm{d}\beta_N} \log f(\bfZ'|\bbeta')\right]$.
				\STATE Set $\beta_L^{(t)}=\mathcal{P}_{\left[0,B_t\right]}\left(\beta_L^{(t-1)} + \Delta \beta_L\right)$ with 
				\STATE $\Delta \beta_L = \delta_t\left[\frac{\textrm{d}}{\textrm{d}\beta_L} \log f(\bfZ|\bbeta') - \frac{\textrm{d}}{\textrm{d}\beta_L} \log f(\bfZ'|\bbeta')\right]$.
				\STATE Set $\beta_0^{(t)}=\mathcal{P}_{\left[0,1\right]}\left(\beta_0^{(t-1)} + \Delta \beta_0\right)$ with 
				\STATE $\Delta \beta_0 = \delta_t\left[\frac{\textrm{d}}{\textrm{d}\beta_0} \log f(\bfZ|\bbeta') - \frac{\textrm{d}}{\textrm{d}\beta_0} \log f(\bfZ'|\bbeta')\right]$.
				\ELSE
				\STATE Set $\bbeta'^{(t)}=\bbeta'^{(t-1)}$
				\ENDIF
        \STATE Set $t = t+1$.
\end{algorithmic}
\end{algogo}
Note that in Algorithm \ref{algo:algo1}, $\mathcal{P}_{\left[0,B_t\right]}(\cdot)$ denotes the projection onto $\left[0,B_t\right]$.
In this paper, we propose an MCMC method that sequentially updates $\bfZ,\MATmat,\MATabond,\bfX,\bsigma^2$ and $s^2$ at each sampler iteration. However, some of these variables, such as $(\bfZ,\bfX)$,$(\MATmat,\bfX)$ or $(\MATabond,\bfX)$ could be updated simultaneously or could be re-sampled within a given sampler iteration, thus improving the mixing properties of the proposed sampling scheme, c.f. \cite{Newstadt2014ssp}.
\section{Generalization to Robust Bayesian Principal Component Analysis}
\label{sec:RBPCA}
As mentioned previously, one of the main contributions of this paper is the introduction of a 3D Ising model to model the possible correlation between the outlier locations. This outlier model has been applied to linear SU in Section \ref{sec:bayesian}. However, this model can be applied to many other blind source separation (subspace identification) and inverse problems (nonlinear spectral unmixing) where the observations are corrupted by additive Gaussian noise and outliers. In this section, we discuss the generalization of the robust Bayesian principal component analysis model studied in \cite{Ding2011,Newstadt2014ssp}.
As in \cite{Ding2011}, we consider the following observation model (expressed in matrix form)
\begin{eqnarray}
\MATpix = \bfL + \bfR + \bfE
\end{eqnarray}
where $\bfR$ (resp. $\bfE$) are $L \times N$ matrices representing the outliers (resp. the additive Gaussian noise) and $\bfL=\bfD \bLambda \bfW\transp$ corresponds to a low rank matrix. 
The $L \times R$ matrices $\bfD$ and $\bfW$
are matrices of the left- and right-singular vectors, respectively, and
$\bLambda=\textrm{diag}\left([\lambda_1,\ldots,\lambda_R]\right)$
is a diagonal matrix consisting of
the singular values $\{\lambda_r\}_{r=1,\ldots,R}$. Following the model considered in \cite{Ding2011,Newstadt2014ssp}, we can set $\lambda_r=\zeta_r \eta_r$ where $\zeta_r \in \{0,1\}$, which allows estimation of the dimension of the principal subspace. Since the columns of $\bfE$ are i.i.d., i.e., 
\begin{eqnarray}
\Vnoise_n \sim \Ndistr{\Vnoise_n;\Vzero_{\nbband},\bSigma_0},\quad \forall n,
\end{eqnarray}
we can assign appropriate prior distributions to $\bsigma^2,\bfD,\bfW,\{\zeta_r\}$ and $\{\eta_r\}$, and consider the Bayesian model proposed in Section \ref{sec:bayesian} for the outlier matrix $\bfR$ (i.e., Eqs. \eqref{eq:outliers_decomp}-\eqref{eq:prior_s2}). The resulting \emph{robust Bayesian principal component analysis model} can then be used to estimate the data principal subspace in the presence of non-i.i.d. additive Gaussian noise and possibly correlated outliers. As the posterior distribution of the anomalous outliers can also be estimated, the model also allows the outlier matrix to be estimated. The sampling strategy associated with Bayesian model is similar to those proposed in \cite{Ding2011,Newstadt2014ssp} and that presented in Section \ref{sec:Gibbs} and is not further developed here due to space constraints.
\vspace{-0.3cm}
\section{Simulations using synthetic data}
\label{sec:simu}
\subsection{Linear mixtures corrupted by outliers}
\begin{figure}[h!]
  \centering
  \includegraphics[width=\columnwidth]{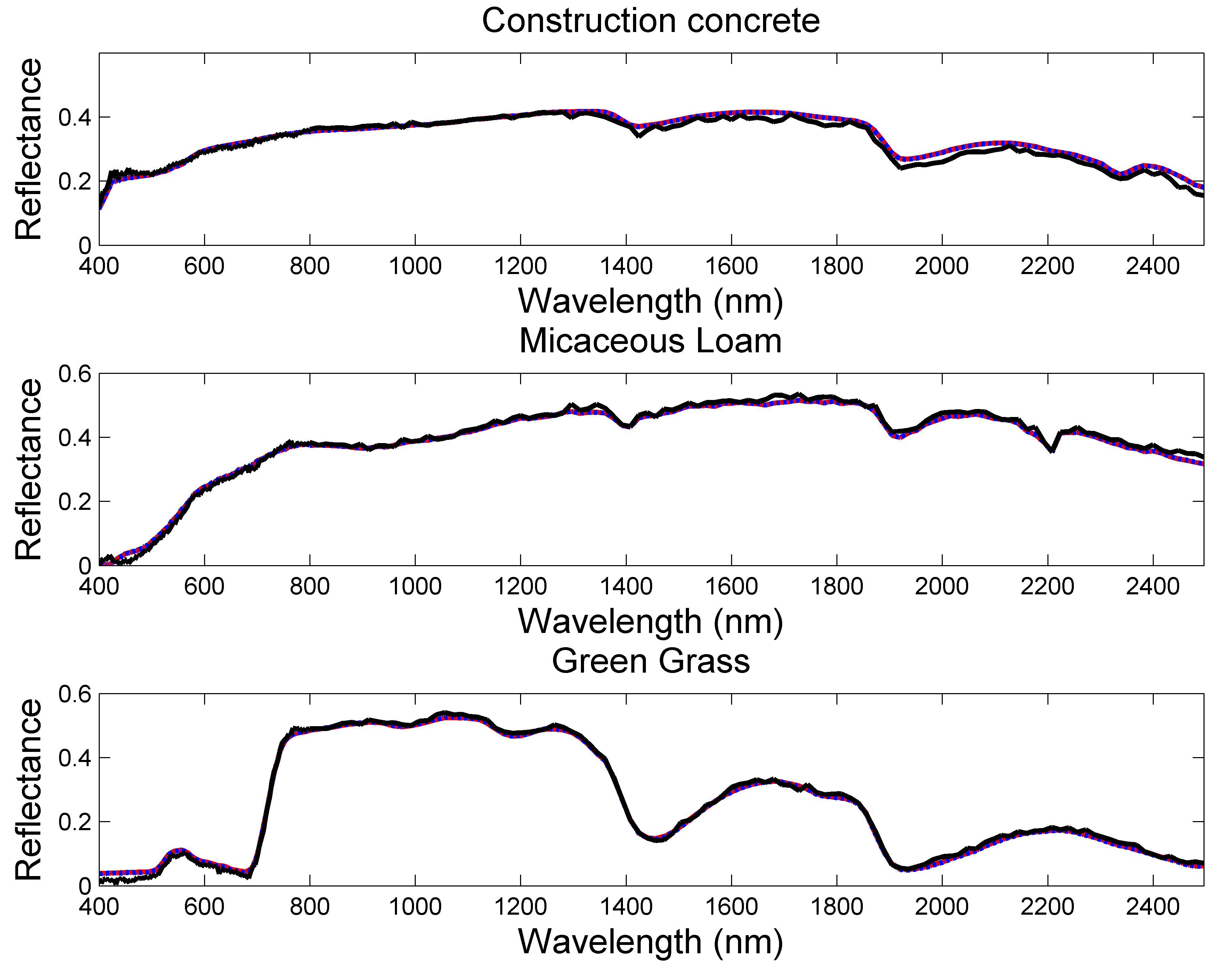}
  \caption{Actual endmembers (red lines) used to generate the synthetic images and endmembers estimated by VCA (black lines) and RBLU (dashed blue lines) for $I_2$.}
  \label{fig:endmembers}
\end{figure}

The performance of the proposed method, referred to as ``RBLU'' (Robust Bayesian linear unmixing), is investigated on two synthetic $60 \times 60$ pixel hyperspectral images composed of
$R=3$ endmembers and observed at $L=207$ different spectral bands (see Fig. \ref{fig:endmembers}). The abundances of the two images are uniformly distributed in the simplex, defined by the positivity and sum-to-one constraints, and the noise variances set to $\sigma_{\ell}^2=10^{-4},\forall \ell$, corresponding to an average SNR of $30$dB without any anomaly addition. The first image $I_1$ does not contain outliers whereas the parameter $s^2$ controlling the outlier power has been set to $s^2=0.1$ for the second image $I_2$. The label matrix of $I_2$ has been generated using \eqref{eq:Potts} with $\bbeta=[0.25;0.25;0.55]\transp$ which leads to approximately $10\%$ of actual outliers in $\bfR$. The proposed method has been applied to the images with $N_{\textrm{MC}}=1000$ iterations (including $N_{\textrm{bi}}=300$). The endmember matrix was initialized using VCA \cite{Nascimento2005} and the abundance matrix was initialized using FCLS \cite{Heinz2001}. The combination VCA-FCLS is also used for performance comparison.  

The quality of the unmixing procedures can be measured by comparing
the estimated and actual abundance vector using the root normalized
mean square error (RNMSE) defined by
\begin{eqnarray}
\label{eq:RMSE}
    \textrm{RNMSE}= \sqrt{\dfrac{1}{\nbpix \nbmat}\sum_{\nopix=1}^{\nbpix}
    {\norm{\hat{\Vabonds}_{\nopix} - \Vabond{\nopix}}^2}}
\end{eqnarray}
where $\Vabond{\nopix}$ and $\hat{\Vabonds}_{\nopix}$ are the actual
and estimated abundance vectors for the $\nopix$th pixel of the
image.
The quality of endmember estimation is evaluated by the
spectral angle mapper (SAM) defined as
\begin{eqnarray}
\label{eq:SAM}
    \textrm{SAM}= \textrm{arccos}\left(\dfrac{\left \langle \hat{\bfm}_r, \Vmat{r}\right \rangle}{\norm{\hat{\bfm}_r} \norm{\Vmat{r}}} \right)
\end{eqnarray}
where $\Vmat{r}$ is the $r$th actual endmember and $\hat{\bfm}_r$ its
estimate. The smaller the value of $|\textrm{SAM}|$, the closer the estimated
endmembers to their actual values.

Table \ref{tab:perf_synth} compares the performance of the proposed method and the VCA-FCLS unmixing strategy and shows that the proposed methods outperforms VCA-FCLS in terms of abundance and endmember estimation. 
Moreover, the confusion matrix of the proposed outlier detection method in Table \ref{tab:confusion} illustrates the ability of the method to identify the corrupted data. 

\begin{table}[h!]
\renewcommand{\arraystretch}{1.2}
\begin{footnotesize}
\begin{center}
\begin{tabular}{|c|c|c|c|c|c|}
\cline{4-6}
\multicolumn{3}{c|}{} & RBLU & VCA-FCLS & o-FCLS\\
\hline
\multirow{4}{*}{$I_1$} & \multirow{3}{*}{SAM ($\times 10^{-2}$)} & $\Vmat{1}$ & \textbf{0.21} & 0.68 & -\\
\cline{3-6}
                   &    & $\Vmat{2}$ & \textbf{0.17} & 0.92 & -\\
\cline{3-6}
                    &   & $\Vmat{3}$ & \textbf{0.26} & 1.96 & -\\
\cline{2-6}
&\multicolumn{2}{|c|}{RNMSE ($\times 10^{-2}$)} & 0.68 & 1.60 & \textbf{0.67}\\
\hline
\hline
\multirow{4}{*}{$I_2$} & \multirow{3}{*}{SAM ($\times 10^{-2}$)} & $\Vmat{1}$ & \textbf{0.19} & 4.03 & -\\
\cline{3-6}
                   &    & $\Vmat{2}$ & \textbf{0.20} & 3.08 & -\\
\cline{3-6}
                    &   & $\Vmat{3}$ & \textbf{0.29} & 4.26 & -\\
\cline{2-6}
&\multicolumn{2}{|c|}{RNMSE ($\times 10^{-2}$)} & \textbf{0.74} & 8.27 & 6.95\\
\hline
\end{tabular}
\caption{Estimation performance.\label{tab:perf_synth}}
\end{center}
\end{footnotesize}
\vspace{-0.4cm}
\end{table}

\begin{table}[h!]
\renewcommand{\arraystretch}{1.2}
\begin{footnotesize}
\begin{center}
\begin{tabular}{|c|c|c|c|}
\cline{2-4}
\multicolumn{1}{c|}{} & $z=0$ & $z=1$ & Total\\
\hline
$\hat{z}=0$ &  652724 & 7190 & 659914\\
\hline
$\hat{z}=1$ & 789 & 84497 & 85286\\
\hline
Total  & 653513 & 91687 & 745200\\
\hline
\end{tabular}
\vspace{0.4cm}
\caption{Outlier detection ($I_2$): confusion matrix.\label{tab:confusion}}
\end{center}
\end{footnotesize}
\vspace{-0.4cm}
\end{table}

\begin{table}[h!]
\renewcommand{\arraystretch}{1.2}
\begin{footnotesize}
\begin{center}
\begin{tabular}{|c|c|c|c|c|c|}
\hline
Noise & Outlier& \multicolumn{2}{|c|}{$s^2=0.01$} & \multicolumn{2}{|c|}{$s^2=0.1$}\\
\cline{3-6}
	variance & fraction & RBLU & o-FCLS & RBLU & o-FCLS\\
\hline
\multirow{3}{*}{$\sigma^2=10^{-4}$} & $10\%$ &  $0.77$ & $2.00$ & $\textbf{0.74}$ & $6.95$\\
\cline{2-6}
 & $20\%$ &  $0.80$ & $3.06$ & $\textbf{0.77}$ & $8.89$\\
\cline{2-6}
 & $30\%$ &  $\textbf{0.86}$ & $3.66$ & $0.87$ & $10.34$\\
\hline
\multirow{3}{*}{$\sigma^2=10^{-3}$} & $10\%$ &  $\textbf{2.19}$ & $2.77$ & $2.21$ & $6.00$\\
\cline{2-6}
 & $20\%$ &  $\textbf{2.36}$ & $3.65$ & $2.39$ & $8.90$\\
\cline{2-6}
 & $30\%$ &  $\textbf{2.59}$ & $4.09$ & $2.62$ & $10.57$\\
\hline
\end{tabular}
\vspace{0.4cm}
\caption{Abundance RNMSE ($\times 10^{-2}$) for different outlier energies and proportions.\label{tab:abond_estim}}
\end{center}
\end{footnotesize}
\vspace{-0.4cm}
\end{table}

Table \ref{tab:abond_estim} compares the abundance estimation performance of RBLU to o-FCLS (which assumes perfectly known endmembers) for different outlier corruption scenarios (proportions and variances) and two noise settings, $\sigma^2=10^{-4}$ and $\sigma^2=10^{-3}$ (which correspond to SNR of approximately $30$dB and $20$dB, respectively, when considering data without anomalies). This table shows a general performance degradation of the algorithms when the number of outliers increases. However, although RBLU also estimates the endmembers (jointly with the abundances), the performance degradation is less severe for RBLU than for o-FCLS by virtue of RBLU's outlier detection ability. It is interesting to note that RBLU is also less sensitive than o-FCLS to variations in the outlier variance (o-FCLS abundance estimation performance decreases when the outlier variance increases). In addition, as the proportion of outliers increases, sparsity ceases to be a reliable discriminant and it becomes increasingly difficult to detect the outlier samples. Similarly, if the variances of the noise and the anomalies are similar RBLU will not be able to detect the potential anomalies, confounding the outliers with a fictitious Gaussian noise having larger variance.

For a $64$-bit Matlab R2014b implementation on a $3$GHz Intel Xeon quad-core workstation, RBLU required $30$min to analyse each image composed of $60 \times 60$ pixels ($3$s for VCA-FCLS). Although RBLU can provide significantly improved endmember and abundance estimates, it has a higher computational cost than VCA-FCLS.

Next we discuss the unmixing problem when both linearly and nonlinearly mixed pixels are present in the image.

\subsection{Unmixing of linear and nonlinear mixtures}
\begin{figure}[h!]
  \centering
  \includegraphics[width=\columnwidth]{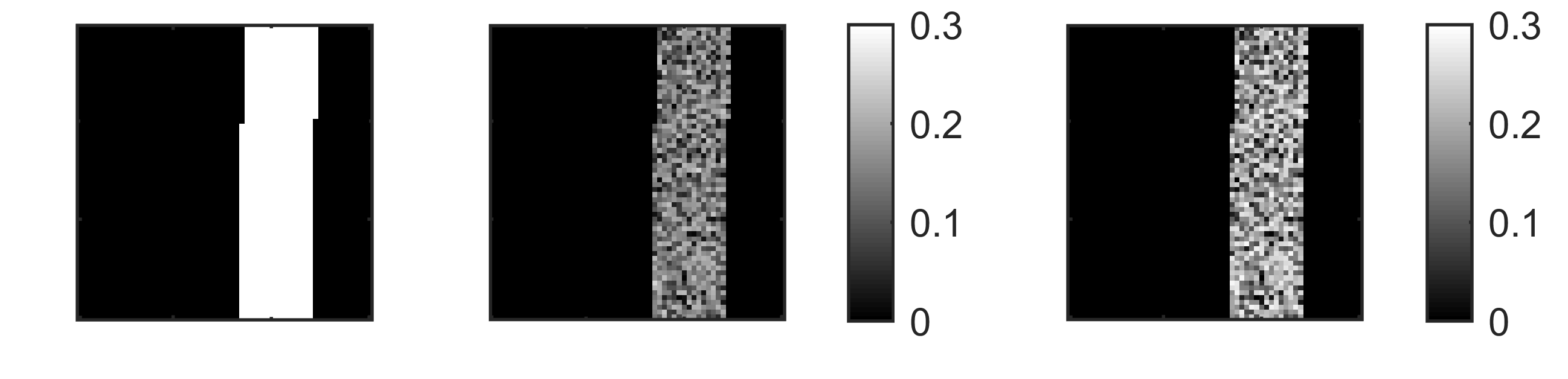}
  \caption{Left: Actual location of the linearly (black pixels) and nonlinearly (white pixels) mixed pixels. Actual (middle) and estimated (right) anomaly energy in each pixel of the synthetic image composed of linearly and nonlinearly mixed pixels.}
  \label{fig:outliers_nonlin}
\end{figure}

As mentioned previously, many nonlinear models of the spectral unmixing literature can be expressed as \eqref{eq:NLM0}. This is the case, for instance, for polynomial models \cite{Somers2009,Nascimento2009,Halimi2010} introduced to handle multiple scattering effects. The proposed model \eqref{eq:NLM0} seems particularly well adapted to detecting nonlinear effects that are often spatially localized, e.g., regions with significant relief variations, and/or spectrally concentrated such as nonlinear interactions (if present) varying smoothly over wavelengths. 
To evaluate the performance of the RBLU algorithm in terms of endmember estimation, abundance estimation and nonlinearity detection, we synthesised a $60 \times 60$ pixel image composed of the three endmembers considered in the previous paragraph. Most of the pixels ($75\%$) were generated according to the classical LMM while the remaining $25\%$ were generated according to a bilinear model, namely, the 
generalized bilinear model \cite{Halimi2010}
\begin{eqnarray}
\label{eq:GBM}
 \Vpix{n} &=& \MATmat \Vabond{n}\nonumber\\
& + &\sum_{i=1}^{R-1}\sum_{j=i+1}^R \gamma_{i,j,n}\abond{i}{n}\abond{j}{n}\Vmat{i}\odot\Vmat{j} + \Vnoise_{n},
\end{eqnarray}
where $0\leq \gamma_{i,j,n} \leq 1$ characterizes the level of interaction between the endmembers $\Vmat{i}$ and $\Vmat{j}$ in the $n$th pixel. This choice of nonlinear model is motivated by the fact the polynomial (and in particular bilinear) models, introduced to account for multiple scattering effects, have demonstrated improved performance for certain types of urban and vegetated areas \cite{Somers2009,Nascimento2009,Halimi2010}.
The abundances of each pixel (linearly or nonlinearly) mixed were uniformly drawn from the simplex defined by the abundance positivity and sum-to-one constraints. The nonlinearity parameters in \eqref{eq:GBM} were fixed to $\gamma_{i,j,n}=1$, which corresponds to the choice made in Fan's model \cite{Fan2009}. All pixels have been corrupted by a zero-mean additive Gaussian noise i.i.d, with variance $\sigma^2$, corresponding to an average SNR of $28$dB. Note that although the generated abundance vectors are spatially independent, the positions of the generated anomalies are spatially correlated due to the spatial organization of the linearly and nonlinearly mixed pixels (see Fig. \ref{fig:outliers_nonlin} (left)). The RBLU algorithm was applied to the data using $N_{\textrm{MC}}=2000$ iterations (including $N_{\textrm{bi}}=500$). 

Table \ref{tab:perf_nonlin} compares the estimation performance of RBLU to the results obtained with 1) BLU \cite{Dobigeon2009}, 2) VCA+FCLS \cite{Nascimento2005,Heinz2001}, 3) NfindR + FCLS \cite{Winter1999,Heinz2001}, 4)o-FCLS \cite{Heinz2001}, 5) NfindR \cite{Winter1999} followed by the gradient-based inversion step proposed in \cite{Halimi2010} based on the GBM \eqref{eq:GBM} (referred to as ``NfindR + GBM'' in the table), 6) VCA \cite{Nascimento2005} followed by the gradient-based inversion step proposed in \cite{Halimi2010} based on the GBM \eqref{eq:GBM} (referred to as ``VCA + GBM'' in the table), and 7) the GBM method \cite{Halimi2010} applied to the data assuming the endmembers are known (referred to as ``o-GBM'' in the table for oracle-GBM). Due to its anomaly detection ability, RBLU generally provides better abundance estimates than LMM-based methods. Although RBLU does not rely on the GBM, its provides better abundance estimates than BLU, in particular for nonlinearly mixed pixels. For this data set containing pure pixels, N-FindR and VCA can provide better endmember estimates than RBLU, which would be different in the absence of pure pixels. Indeed, in a similar manner to BLU, the proposed RBLU algorithm does not require the pure pixel assumption. However, due to space constraints, simulations conducted on synthetic data which do not contain pure pixels are not presented in this paper which focuses on the anomaly detection capability of RBLU (the interested reader is invited to consult \cite{Altmann_RBLU_arxiv2015} for additional results on data which do not contain pure pixels). We discuss this point in the next section in the context of the RBLU analysis of real hyperspectral images.

 \begin{table}[h!]
\renewcommand{\arraystretch}{1.2}
\begin{footnotesize}
\begin{center}
\begin{tabular}{|c|c|c|c|c|}
\cline{2-5}
\multicolumn{1}{c|}{} & RNMSE &  \multicolumn{3}{|c|}{SAM $(\times 10^{-2})$}\\
\cline{3-5}
\multicolumn{1}{c|}{}& $(\times 10^{-2})$ & $\Vmat{1}$& $\Vmat{2}$& $\Vmat{3}$\\
\hline
o-FCLS & $4.44$ & $-$ & $-$ & $-$\\
\hline
VCA+FCLS & $4.46$ & $\textbf{0.35}$ & $\textbf{0.55}$ & $0.90$\\
\hline
NfindR + FCLS & $5.96$ & $0.65$ & $3.09$ & $0.80$\\
\hline
o-GBM & $2.28$ & $-$ & $-$ & $-$\\
\hline
VCA+GBM & $\textbf{2.15}$ & $\textbf{0.35}$ & $\textbf{0.55}$ & $0.90$\\
\hline
NfindR + GBM & $2.26$ & $0.65$ & $3.09$ & $0.80$\\
\hline
BLU & $4.03$ & $1.76$ & $1.94$ & $1.33$\\
\hline
RBLU & $3.81$ & $2.79$ & $3.56$ & $\textbf{0.59}$\\
\hline
\end{tabular}
\vspace{0.4cm}
\caption{Abundance and endmember estimation for the synthetic image composed of linearly and nonlinearly mixed pixels.\label{tab:perf_nonlin}}
\end{center}
\end{footnotesize}
\vspace{-0.4cm}
\end{table}

Fig. \ref{fig:outliers_nonlin} (middle and right) shows the actual and estimated outlier energy in each pixel (i.e., $\norm{\hat{\bfr}_{n}}_2^2$) and illustrates the ability of RBLU to detect bilinear mixtures in an image containing both linear and non-linear mixtures.

 \begin{table}[h!]
\renewcommand{\arraystretch}{1.2}
\begin{footnotesize}
\begin{center}
\begin{tabular}{|c|c|}
\hline
o-FCLS & $1$ \\
\hline
VCA+FCLS & $3$ \\
\hline
NfindR + FCLS & $3$ \\
\hline
o-GBM & $ 350$ \\
\hline
VCA+GBM & $ 354$ \\
\hline
NfindR + GBM & $358$ \\
\hline
BLU & $583$ \\
\hline
RBLU & $1526$ \\
\hline
\end{tabular}
\vspace{0.4cm}
\caption{Computational time (in seconds) for the synthetic image composed of linearly and nonlinearly mixed pixels.\label{tab:compute_synth1}}
\end{center}
\end{footnotesize}
\vspace{-0.4cm}
\end{table}

Finally, Table \ref{tab:compute_synth1} compares the execution time required by the different methods to analyse the $3600$ pixels of the image composed of linearly and nonlinearly mixed pixels on the same basis as before. Although the GBM-based method proposed \cite{Halimi2010} takes longer than FCLS, it processes the pixels independently and successively. Its run-time could thus be improved (\emph{e.g.}, using parallelization) and its computational complexity could approach that of FCLS. BLU and RBLU are unsupervised unmixing algorithms, which jointly estimate the endmembers and abundances and are based on MCMC methods, which are more computationally demanding. Although some sampling steps can be performed in parallel, the underlying Gibbs samplers are intrinsically sequential processes that require a sufficient number of iterations to explore the posterior distribution of interest. RBLU is more computationally demanding than BLU as it includes additional sampling steps (labels and anomaly values) and the estimation of the Ising model regularization parameters. However, the results presented in this paper illustrate the benefits of the (structured) anomaly detection ability of RBLU on the endmember and abundance estimation performance.

\vspace{-0.3cm}
\section{Simulations using real hyperspectral data}
\label{sec:simu_real}
\subsection{Moffett data set}
The first real image considered in this section is composed
of $L=189$ spectral bands and was acquired in $1997$ by the
AVIRIS satellite. The acquired image covers a region over Moffett Field, CA. A subimage of size
$50 \times 50$ pixels has been chosen here to evaluate the performance of RBLU. The AVIRIS Moffet Field dataset has been previously used for comparing methods of linear \cite{Besson_IEEE_Trans_SP_2011,Halimi2010,Eches2010a,Eches2010,Dobigeon2009} and nonlinear \cite{Halimi2010} unmixing. The subimage of interest is mainly composed of
water, vegetation, and soil. As in in the previous subsection, the RBLU algorithm was applied with $N_{\textrm{MC}}=2000$ iterations (including $N_{\textrm{bi}}=500$). 

\begin{figure}[h!]
  \centering
  \includegraphics[width=\columnwidth]{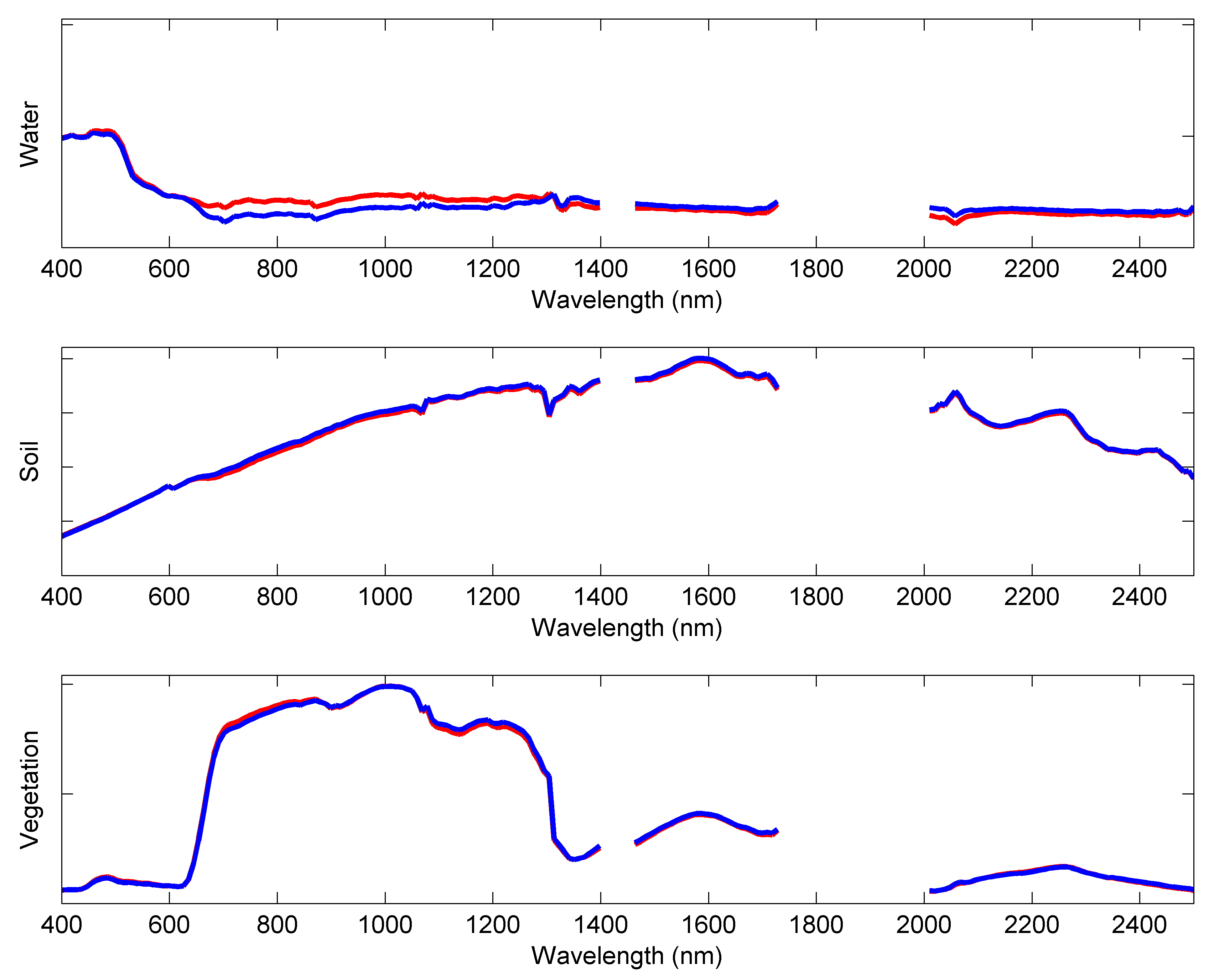}
  \caption{The $R=3$ endmembers extracted from the Moffett image by N-FindR (red) and RBLU (blue).}
  \label{fig:endm_Moffett}
\end{figure}

\begin{figure}[h!]
  \centering
  \includegraphics[width=\columnwidth]{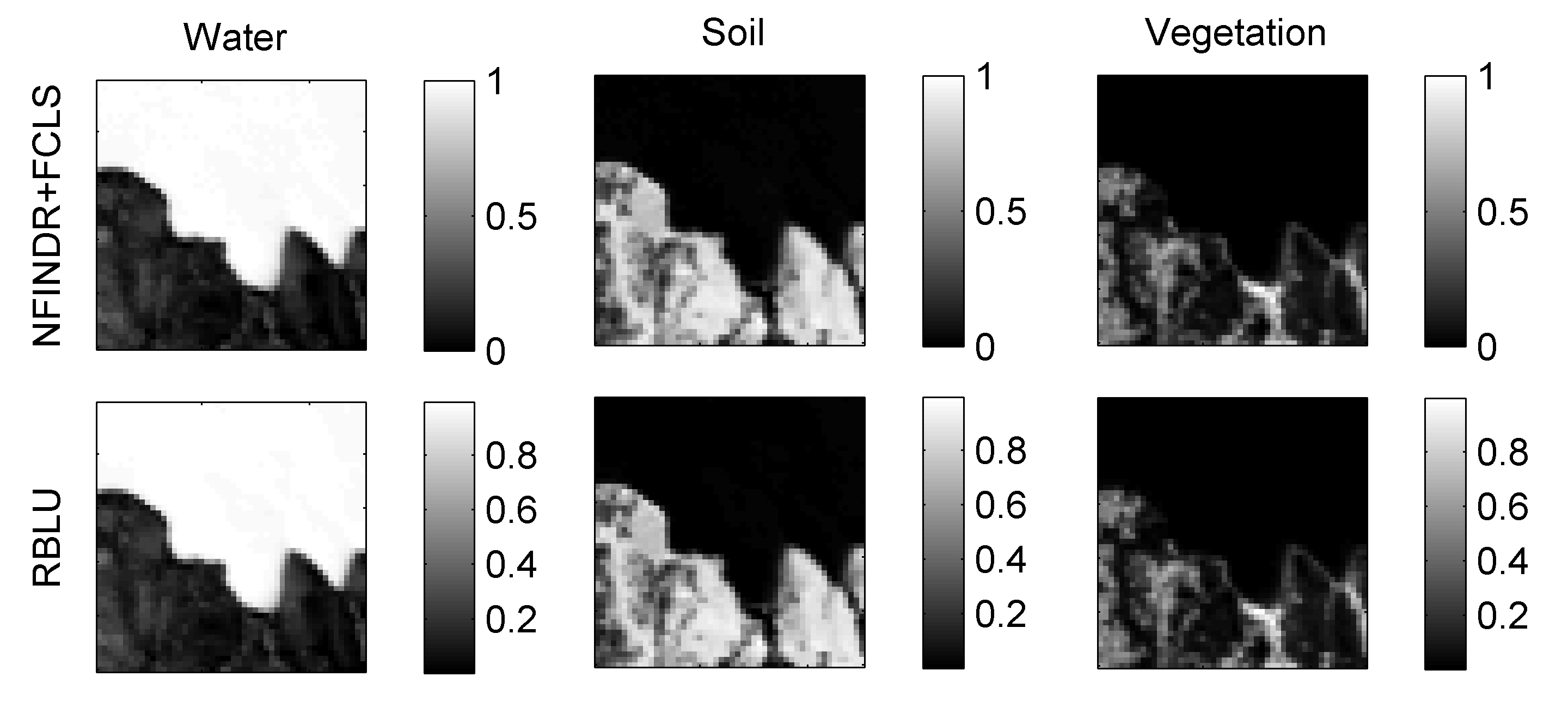}
  \caption{The $R=3$ abundance maps associated with the Moffett image and estimated by FCLS (top) and RBLU (bottom).}
  \label{fig:abund_Moffett}
\end{figure}

\begin{figure}[h!]
  \centering
  \includegraphics[width=\columnwidth]{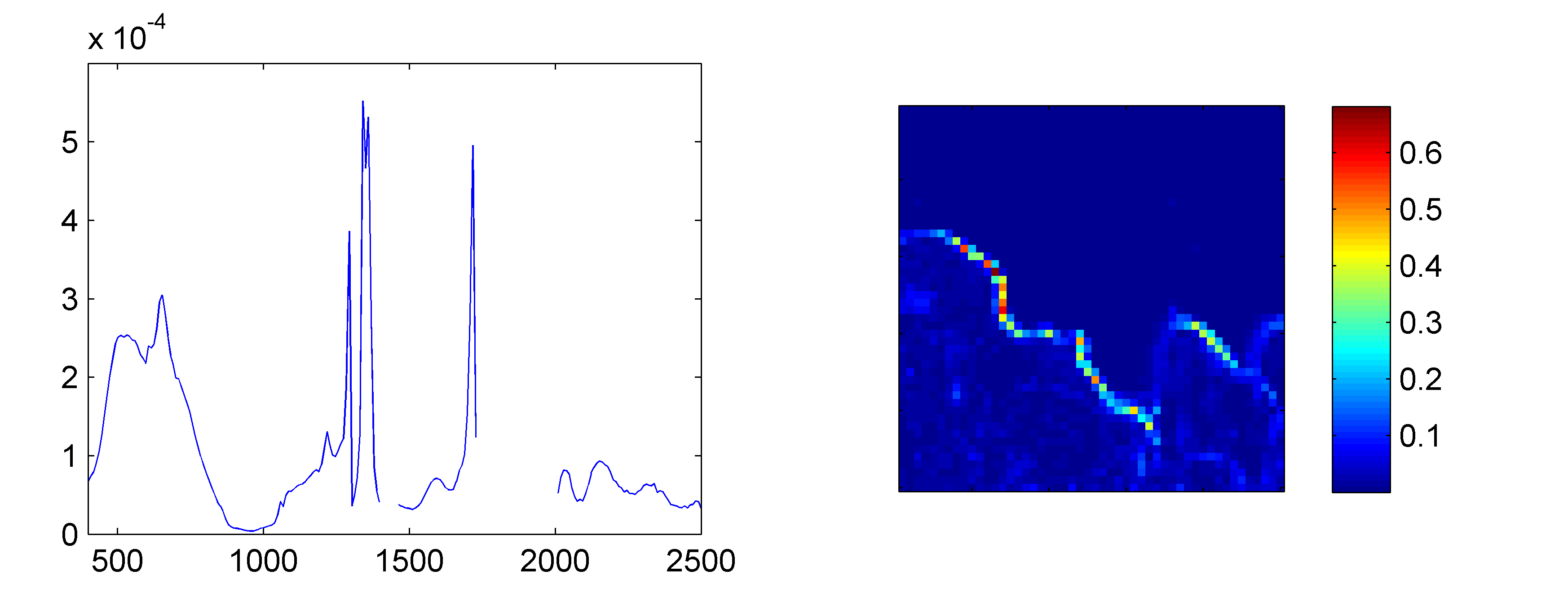}
  \caption{Left: Average spectral energy of the outliers in the Moffett scene estimated by RBLU. Right: Estimated outlier energy $\norm{\hat{\bfr}_{n}}_2^2$ in each pixel of the Moffett image.}
  \label{fig:Outlier_map_Moffett}
\end{figure}

\begin{figure}[h!]
  \centering
  \includegraphics[width=\columnwidth]{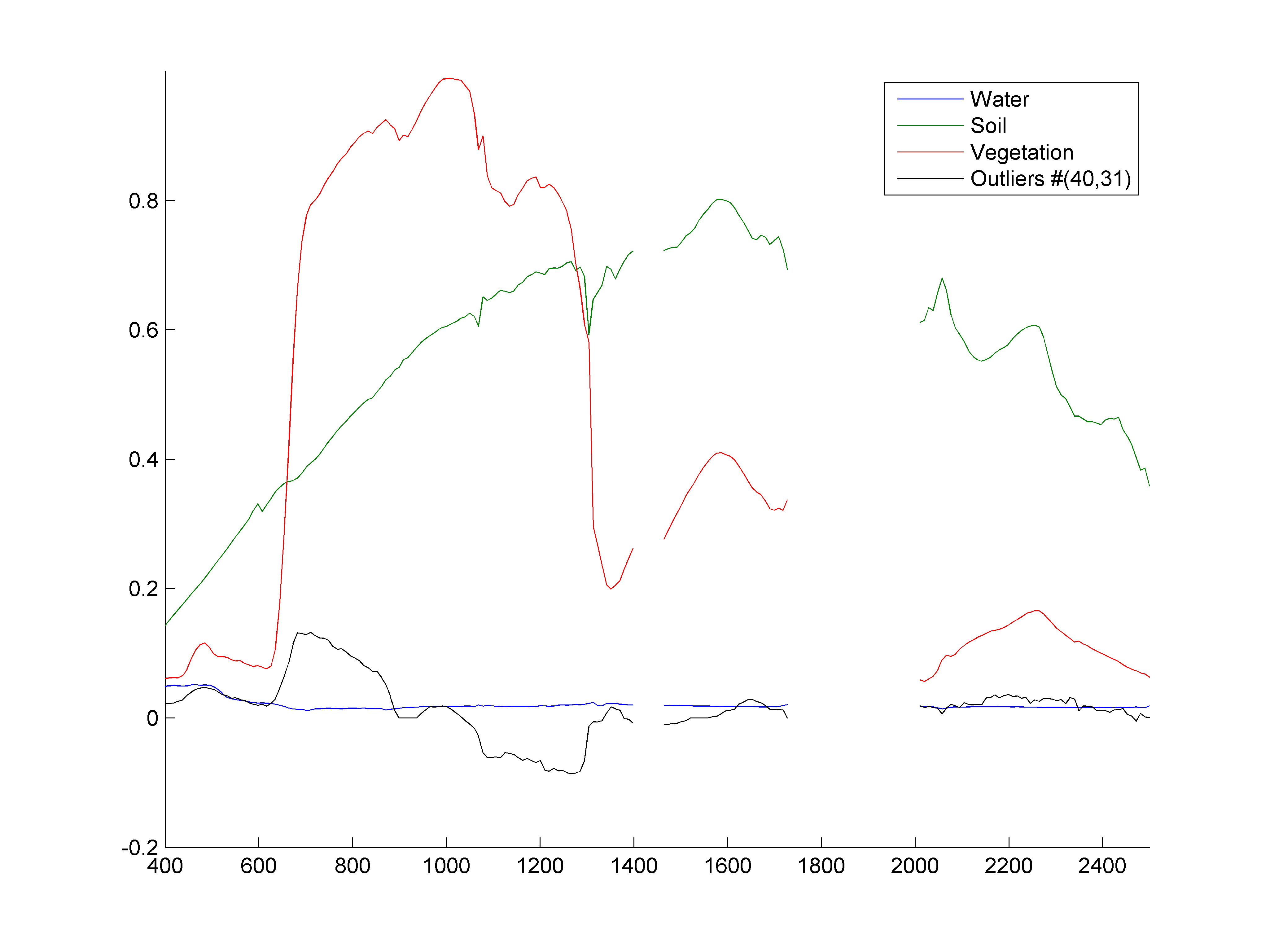}
  \caption{Endmembers and outlier signature of the pixel (40,31) estimated by RBLU for the Moffett image.}
  \label{fig:ex_outlier_Moffett}
\end{figure}

\begin{figure}[h!]
  \centering
  \includegraphics[width=\columnwidth]{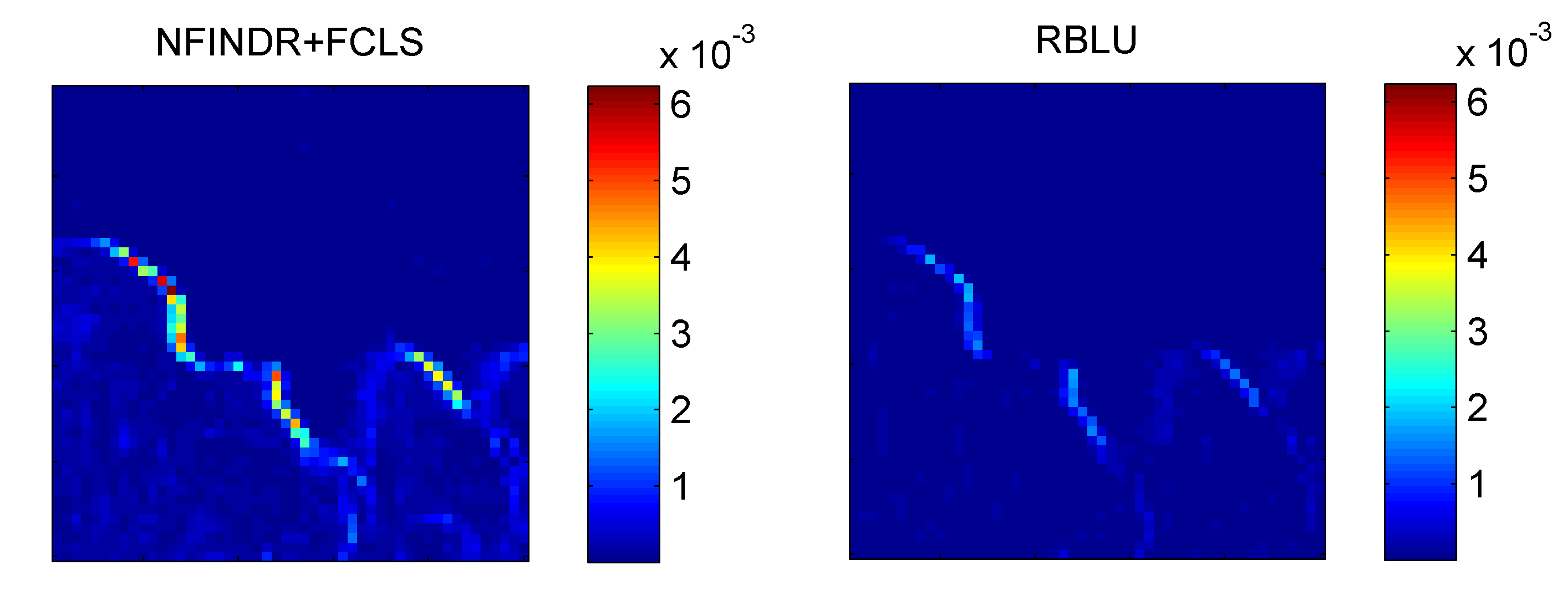}
  \caption{Reconstruction errors of the Moffett image using N-FindR+FCLS (left) and RBLU.}
  \label{fig:reconstruction_Moffett}
\end{figure}

Fig. \ref{fig:endm_Moffett} depicts the endmembers estimated by N-FindR and RBLU. Fig. \ref{fig:abund_Moffett} compares the abundance maps provided by RBLU to those obtained with N-FindR followed by FCLS. These figures show that the endmembers and abundance maps are all in good agreement. In addition to the endmember and abundance estimates, RBLU provides spectral and spatial information about the possible outliers/nonlinearities. Fig. \ref{fig:Outlier_map_Moffett} (left) shows average spectral outlier energy over wavelength for the Moffett subimage. In addition to significant outlier levels near the water absorption bands, around $1400$nm and $1800$nm, RBLU identifies important deviations from the linear mixing model for the spectral bands between $400$nm and $800$nm. Fig. \ref{fig:Outlier_map_Moffett} (right) displays the estimated outlier energy (i.e., $\norm{\hat{\bfr}_{n}}_2^2$) map over all pixels in the Moffett scene and shows that the deviations from the linear mixing model are mainly located in the coastal area, which is in agreement with the results obtained in \cite{Besson_IEEE_Trans_SP_2011,Halimi2010}. Fig. \ref{fig:ex_outlier_Moffett} compares the estimated spectrum of the outliers in the pixel $(40,31)$ (located in the coastal area) to the estimated endmembers. Although RBLU promotes groups of outliers, it does not explicitly enforce spatial nor spectral dependencies for the outlier values. However, the outliers in this pixel seem to be spectrally correlated. These results show that RBLU is able to distinguish structured outliers from Gaussian noise. These results also show that the outlier spectrum and the vegetation spectral signature (in red) seem correlated, especially in the visible spectrum. A possible explanation for this correlation could be local significant changes of vegetation, e.g., chlorophyll and water content, additional vegetation species, multiple scattering effects. Finally, the reconstruction errors obtained by N-FindR+FCLS and RBLU are compared in Fig. \ref{fig:reconstruction_Moffett}. Due to its outlier detection ability, RBLU provides lower reconstruction errors in the coastal area, which is the region where outliers/anomalies are the most dominant.    

\subsection{Villelongue data set}
\begin{figure}[h!]
  \centering
  \includegraphics[width=\columnwidth]{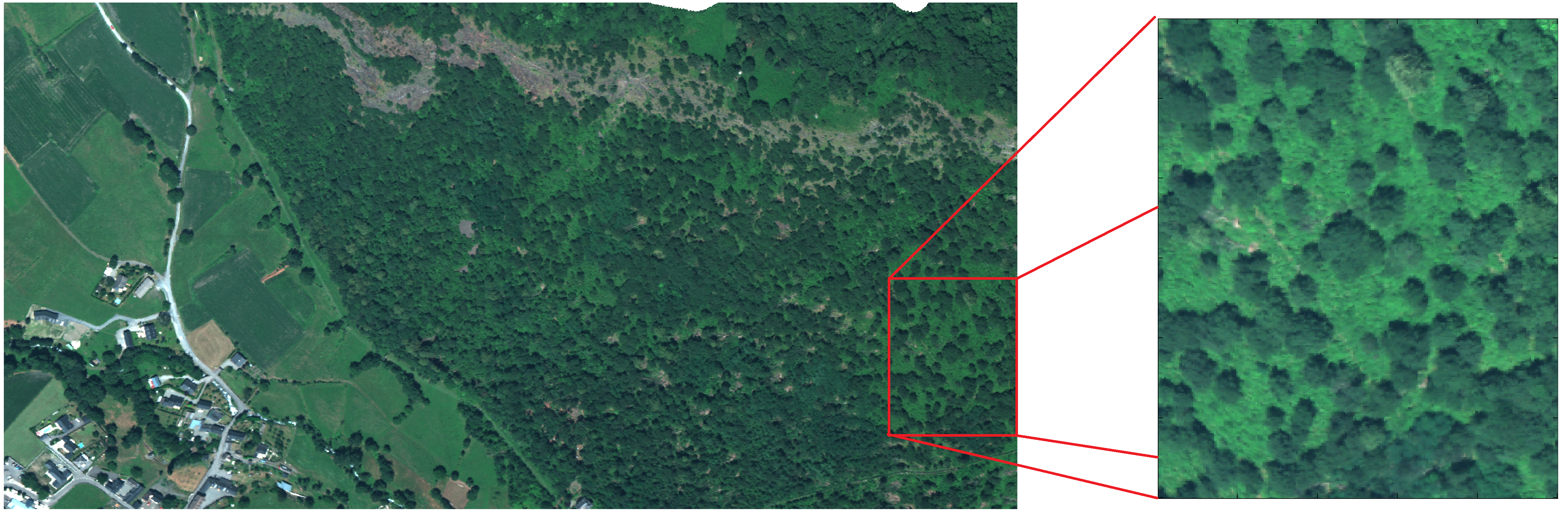}
  \caption{True color image of the Villelongue area (left) and sub-image of interest (right).}
  \label{fig:Madonna_big}
\end{figure}

The second real image considered was acquired in
2010 in the Madonna project and collected by the Hyspex hyperspectral scanner over Villelongue,
France ($00^\circ03'$W and $42^\circ57'$N). $L=
160$ spectral bands were recorded from the visible to near infrared with a spatial
resolution of $0.5$m. This dataset has previously been studied
in \cite{Sheeren2011,Altmann2014a,Altmann2013} and is mainly composed of forested and
urban areas. More details about the data acquisition and pre-processing steps can be found in \cite{Sheeren2011}. A sub-image of size
$300 \times 250$ pixels is chosen here to evaluate the proposed
unmixing procedure and is depicted in Fig. \ref{fig:Madonna_big}. The scene is
composed mainly of trees and grass, resulting in $R=4$ endmembers (soil, grass, trees and shade).

\begin{figure}[h!]
  \centering
  \includegraphics[width=\columnwidth]{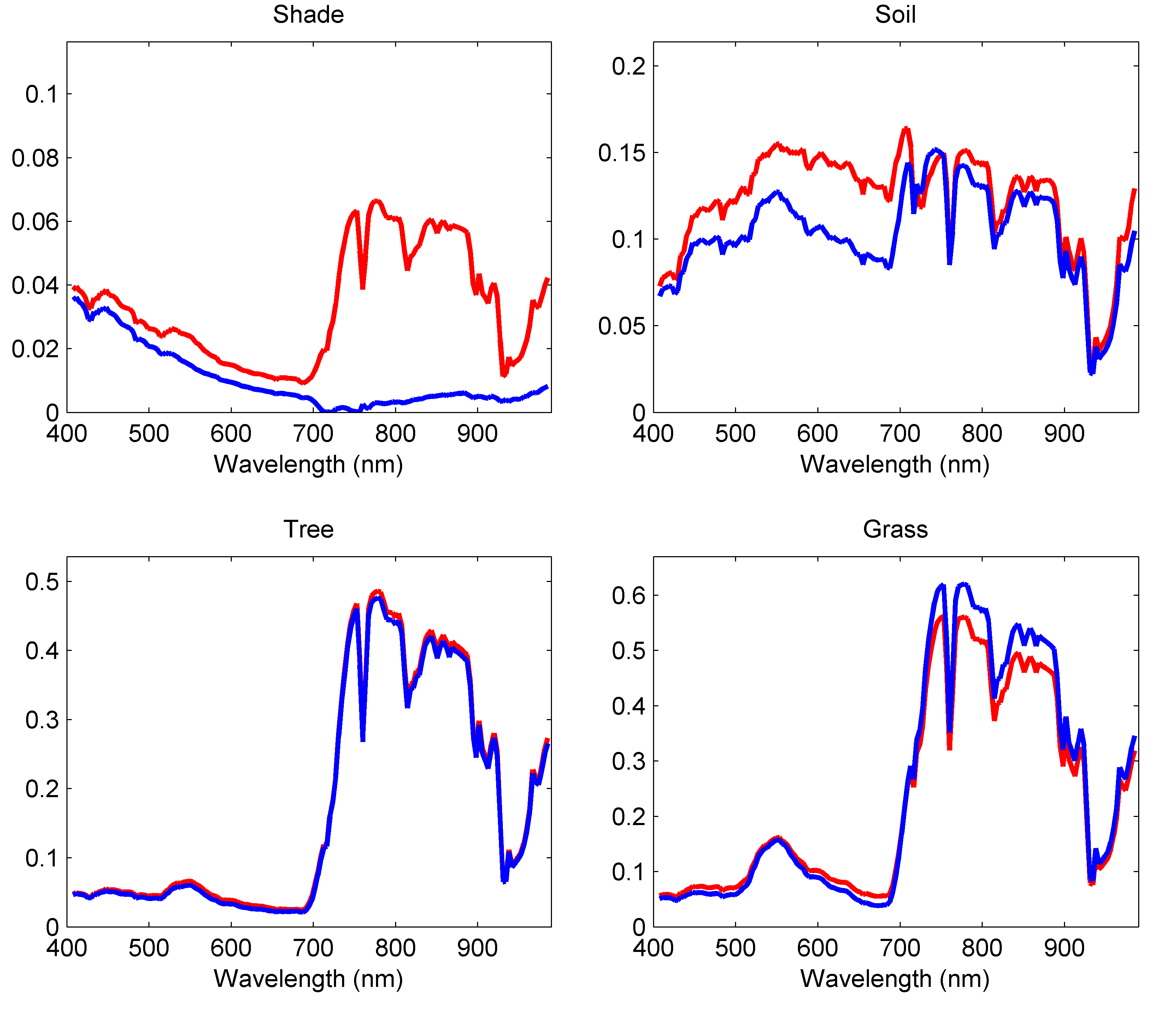}
  \caption{The $R=4$ endmembers extracted from the Villelongue image by N-FindR (red) and RBLU (blue).}
  \label{fig:endm_Madonna}
\end{figure}

\begin{figure}[h!]
  \centering
  \includegraphics[width=\columnwidth]{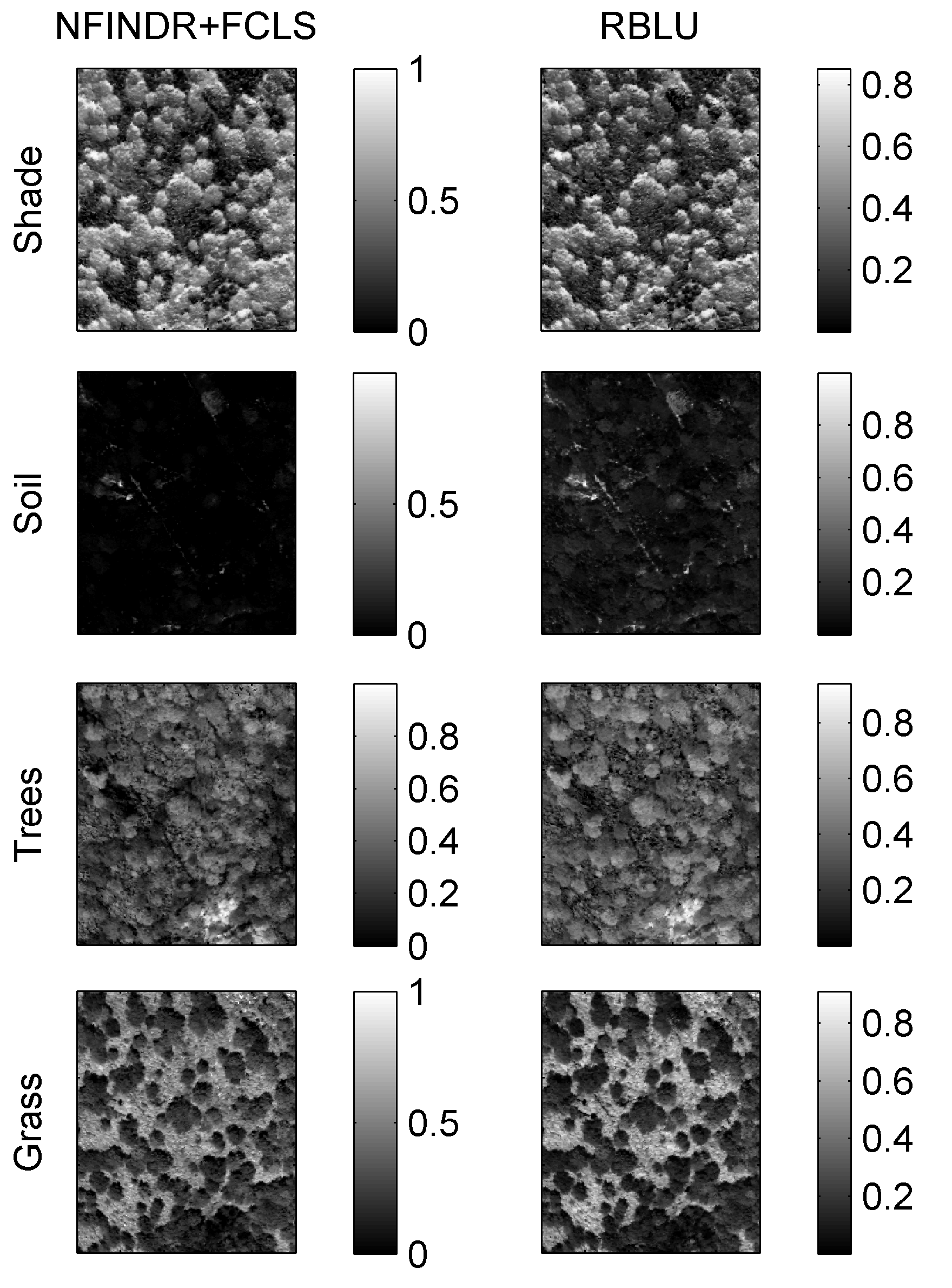}
  \caption{The $R=4$ abundance maps associated with the Villelongue image and estimated by FCLS (right) and RBLU (left).}
  \label{fig:abund_Madonna}
\end{figure}

\begin{figure}
   \begin{minipage}[c]{.48\linewidth}
      \includegraphics[width=\columnwidth]{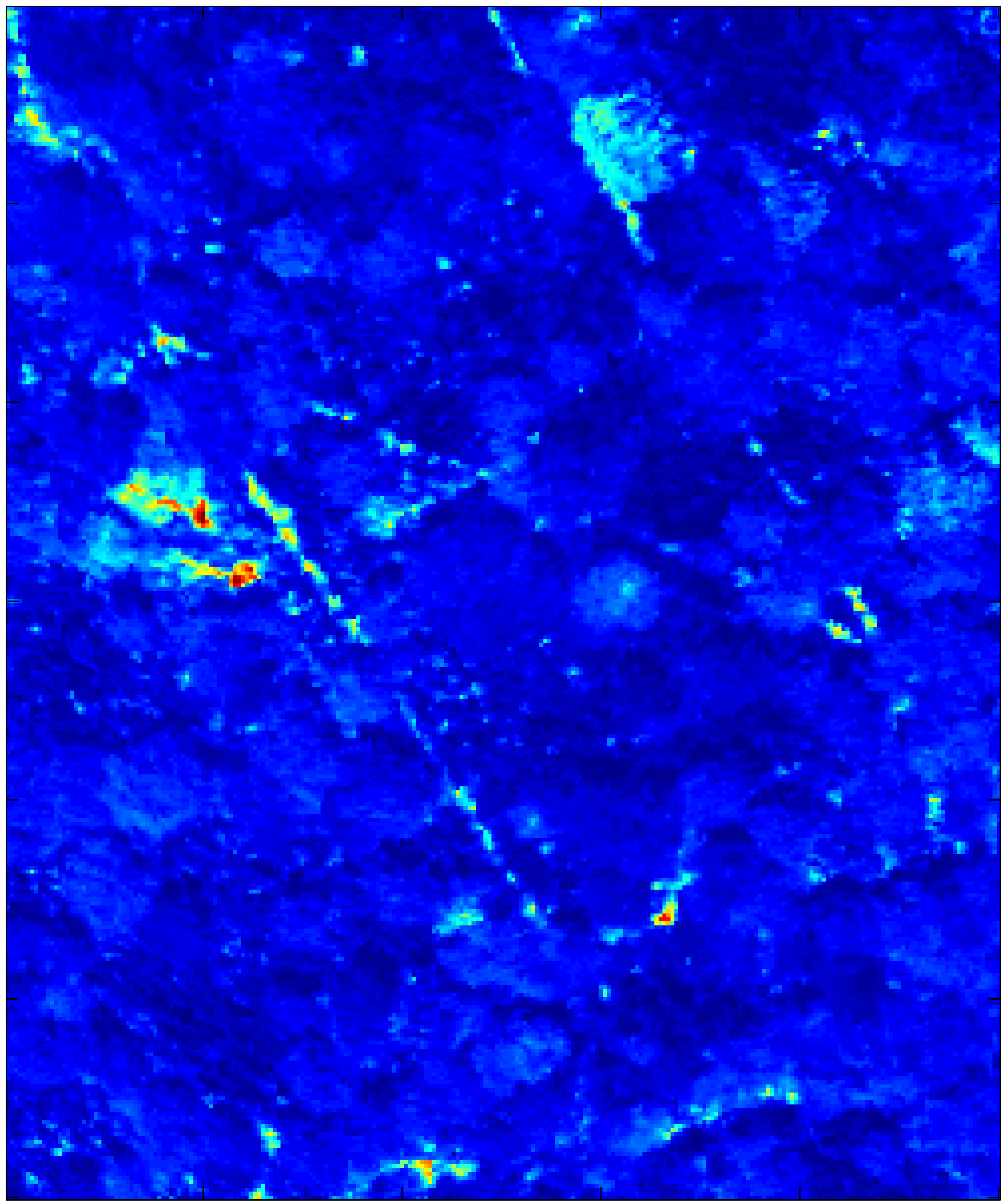}
   \end{minipage} \hfill
   \begin{minipage}[c]{.48\linewidth}
      \includegraphics[width=\columnwidth]{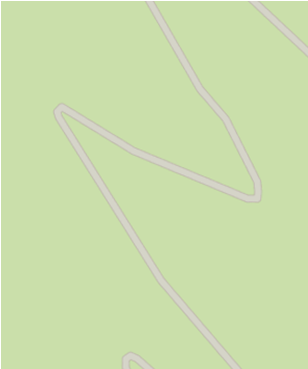}
   \end{minipage}
	\caption{Soil abundance map estimated by RBLU for the Villelongue image (left) and corresponding Google Map image (right) highlighting the presence of a path in the region of interest.}
	\label{fig:Madonna_path}
\end{figure}

Fig. \ref{fig:endm_Madonna} compares the $R=4$ endmembers estimated by N-FindR and RBLU for this second real image. Although it is difficult to objectively assess the performance of the two EEAs for this image, it is interesting to note that the results obtained by the two methods are similar for the tree, soil and grass spectra. The shade signature identified by RBLU has a lower amplitude than the spectrum estimated by N-FindR, which is probably due to the absence of completely shadowed pixels in the image (as discussed above, RBLU does not rely on the pure pixel assumption, in contrast to N-FindR). The two methods however lead to abundance maps (depicted in Fig. \ref{fig:abund_Madonna}) that are in agreement with the true color image in Fig. \ref{fig:Madonna_big}. In particular, the two algorithms are able to identify the path (soil) in the scene. This path is barely visible in Fig. \ref{fig:Madonna_big} but its presence can be confirmed using the Google Map image for this region (see Fig. \ref{fig:Madonna_path}).

\begin{figure}
   \begin{minipage}[c]{.48\linewidth}
      \includegraphics[width=\columnwidth]{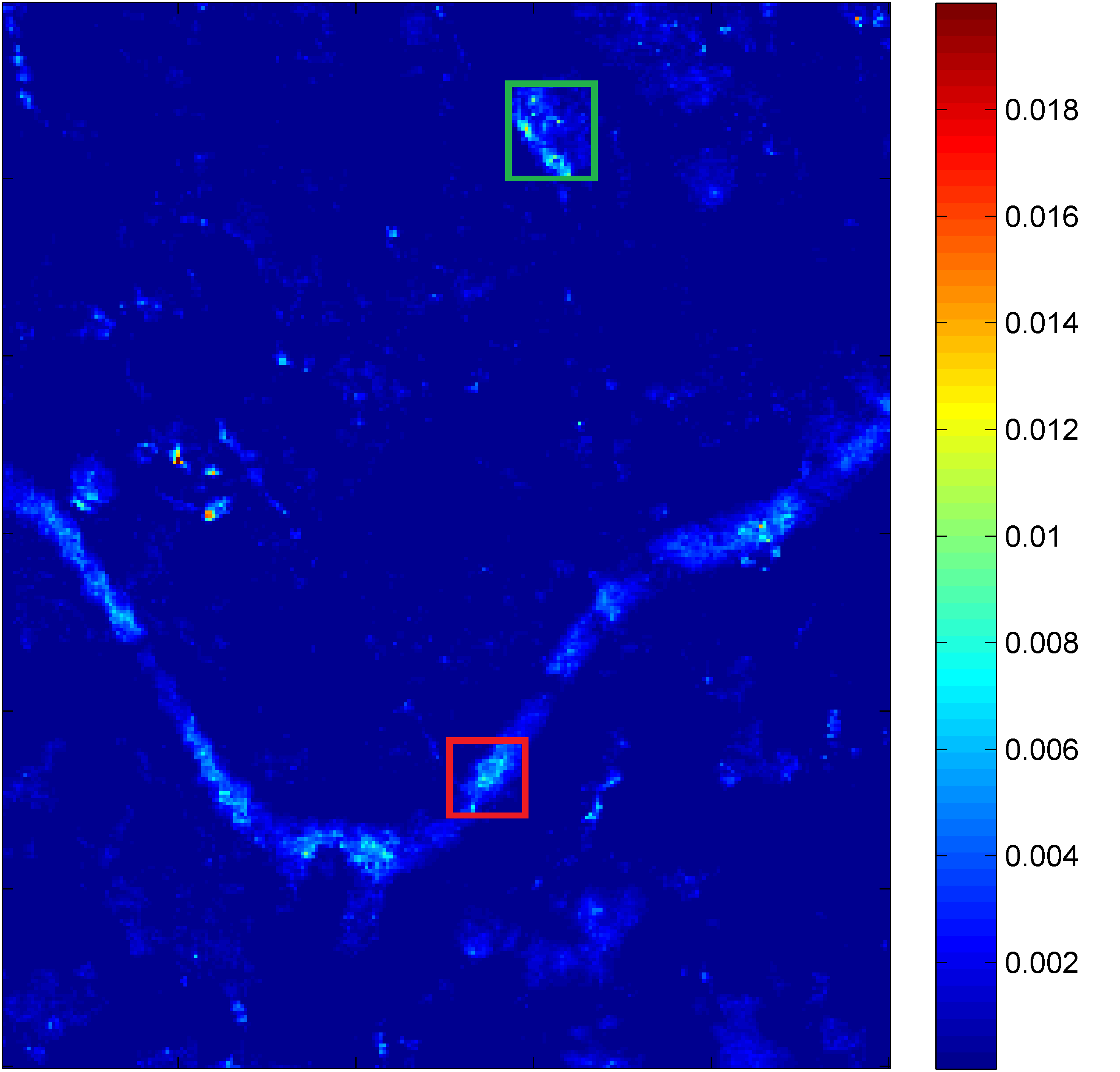}
   \end{minipage} \hfill
   \begin{minipage}[c]{.48\linewidth}
      \includegraphics[width=0.8\columnwidth]{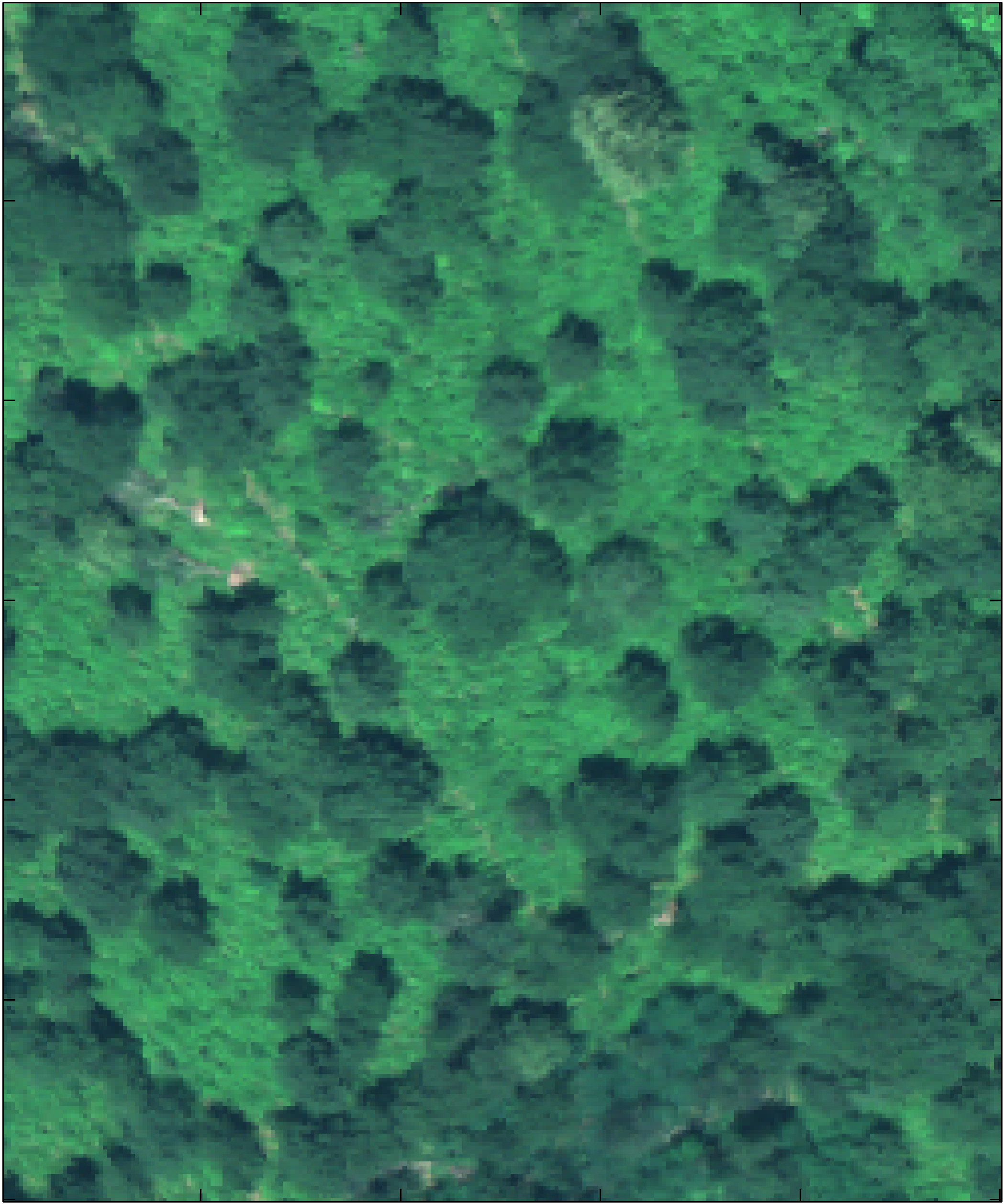}
   \end{minipage}
	\caption{Left: Anomaly energy map estimated by RBLU for the Villelongue scene. Right: sub-image of interest in true colors.}
	\label{fig:Madonna_outliers}
\end{figure}

\begin{figure}[h!]
  \centering
  \includegraphics[width=\columnwidth]{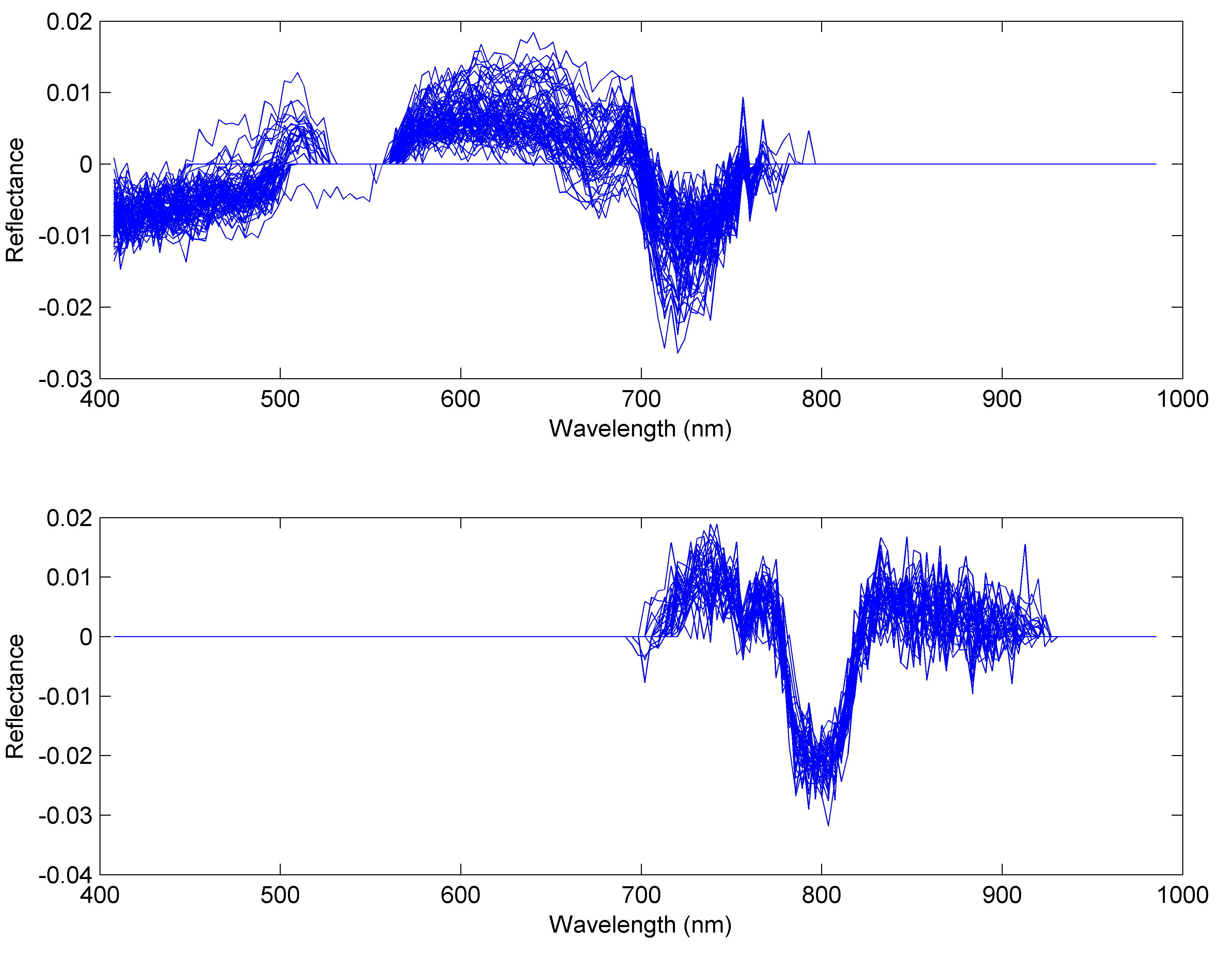}
  \caption{Anomaly spectra in the green (top) and red (bottom) boxes depicted in Fig \ref{fig:Madonna_outliers}.}
  \label{fig:outlier_spectra_Madonna_big}
\end{figure}

Fig. \ref{fig:Madonna_outliers} (left) depicts the anomaly energy map estimated by RBLU over the Villelongue image and highlights two main regions where significant deviations from the linear mixing model occur. The first region, on the top of Fig. \ref{fig:Madonna_outliers} (left) is located where trees are identified and the deviations are likely to be due to the presence of different tree species. Note that this region can also be identified in Fig. \ref{fig:Madonna_outliers} (right) (lighter green region). The second region representing a line in the centre of Fig. \ref{fig:Madonna_outliers} (left) is more difficult to interpret and is barely visible in the true color image. However, it is interesting to note that the anomalies in this region form a stripe whose energy is higher in the pixels composed of grass than it is in those containing trees. Consequently, it is reasonable to postulate that the physical phenomena causing deviations from the linear mixing model occur on the ground or below the surface, i.e., not in the canopy. One possible cause of these spectral signature changes could be the presence of different kinds of surface vegetation. Finally, Fig. \ref{fig:outlier_spectra_Madonna_big} presents examples of anomaly spectra of the first (green box in Fig. \ref{fig:Madonna_outliers}) and second (red box in Fig. \ref{fig:Madonna_outliers}) spatial regions. This figure shows that the deviations from the linear mixing model occur in specific spectral bands and are relatively similar within each spatial region. Moreover, the anomalies are spectrally different in the two regions, which confirms they are probably due to different physical phenomena.

Finally, Table \ref{tab:comput_time_real} compares the Matlab execution time required to analyse the Moffett and Villelongue data using RBLU and N-FindR+FCLS.
\begin{table}[h!]
\renewcommand{\arraystretch}{1.2}
\begin{footnotesize}
\begin{center}
\begin{tabular}{|c|c|c|}
\cline{2-3}
\multicolumn{1}{c|}{} & Moffett & Villelongue\\
\hline
RBLU &  1590 & 18360\\
\hline
N-FindR+FCLS & 3 & 31\\
\hline
\end{tabular}
\vspace{0.4cm}
\caption{Computational time (in seconds) to analyse the real images using RBLU and N-FindR+FCLS.\label{tab:comput_time_real}}
\end{center}
\end{footnotesize}
\vspace{-0.4cm}
\end{table}
\vspace{-0.3cm}
\section{Conclusion}
\label{sec:conclusion}
In this paper, we have proposed a Bayesian algorithm for robust linear spectral unmixing of hyperspectral images that performs joint endmember estimation, abundance estimation and outlier/anomaly detection. Appropriate prior distributions were used to enforce endmember and abundance positivity in addition to abundance sum-to-one constraints. Moreover, a 3D spatial-spectral Ising Markov random field was used to model correlations between outliers. Finally, an adaptive MCMC method was proposed to sample from the resulting posterior distribution in order to estimate the unknown model  parameters. Simulations conducted on synthetic data showed superior performance of the proposed method for linear SU and the detection of outliers in hyperspectral images. The proposed method was also applied to real hyperspectral images and provided interesting results in terms of outlier analysis. Future work might include generalization of the proposed method for non-Gaussian observation noise.

\bibliographystyle{IEEEtran}
\bibliography{biblio}

\begin{thebibliography}{10}
\providecommand{\url}[1]{#1}
\csname url@samestyle\endcsname
\providecommand{\newblock}{\relax}
\providecommand{\bibinfo}[2]{#2}
\providecommand{\BIBentrySTDinterwordspacing}{\spaceskip=0pt\relax}
\providecommand{\BIBentryALTinterwordstretchfactor}{4}
\providecommand{\BIBentryALTinterwordspacing}{\spaceskip=\fontdimen2\font plus
\BIBentryALTinterwordstretchfactor\fontdimen3\font minus
  \fontdimen4\font\relax}
\providecommand{\BIBforeignlanguage}[2]{{%
\expandafter\ifx\csname l@#1\endcsname\relax
\typeout{** WARNING: IEEEtran.bst: No hyphenation pattern has been}%
\typeout{** loaded for the language `#1'. Using the pattern for}%
\typeout{** the default language instead.}%
\else
\language=\csname l@#1\endcsname
\fi
#2}}
\providecommand{\BIBdecl}{\relax}
\BIBdecl

\bibitem{Heinz2001}
D.~C. Heinz and {C.-I Chang}, ``Fully constrained least-squares linear spectral
  mixture analysis method for material quantification in hyperspectral
  imagery,'' \emph{IEEE Trans. Geosci. and Remote Sensing}, vol.~29, no.~3, pp.
  529--545, March 2001.

\bibitem{Bioucas2012}
J.~M. Bioucas-Dias, A.~Plaza, N.~Dobigeon, M.~Parente, Q.~Du, P.~Gader, and
  J.~Chanussot, ``Hyperspectral unmixing overview: Geometrical, statistical,
  and sparse regression-based approaches,'' \emph{IEEE J. Sel. Topics Appl.
  Earth Observations Remote Sensing}, vol.~5, no.~2, pp. 354--379, April 2012.

\bibitem{Dobigeon2013spmag}
N.~Dobigeon, J.-Y. Tourneret, C.~Richard, J.~C.~M. Bermudez, S.~McLaughlin, and
  A.~O. Hero, ``Nonlinear unmixing of hyperspectral images: Models and
  algorithms,'' \emph{IEEE Signal Processing Magazine}, vol.~31, no.~1, pp.
  82--94, Jan. 2014.

\bibitem{Heylen2014}
R.~Heylen, M.~Parente, and P.~Gader, ``A review of nonlinear hyperspectral
  unmixing methods,'' \emph{IEEE J. Sel. Topics in Appl. Earth Observations and
  Remote Sensing}, vol.~7, no.~6, pp. 1844--1868, June 2014.

\bibitem{Hapke1981}
B.~W. Hapke, ``Bidirectional reflectance spectroscopy. {I}. {T}heory,''
  \emph{J. Geophys. Res.}, vol.~86, pp. 3039--�3054, 1981.

\bibitem{Somers2009}
B.~Somers, K.~Cools, S.~Delalieux, J.~Stuckens, D.~V. der Zande, W.~W.
  Verstraeten, and P.~Coppin, ``Nonlinear hyperspectral mixture analysis for
  tree cover estimates in orchards,'' \emph{Remote Sensing of Environment},
  vol. 113, no.~6, pp. 1183--1193, 2009.

\bibitem{Nascimento2009}
J.~M.~P. Nascimento and J.~M. {Bioucas-Dias}, ``Nonlinear mixture model for
  hyperspectral unmixing,'' in \emph{Proc. SPIE Image and Signal Processing for
  Remote Sensing XV}, L.~Bruzzone, C.~Notarnicola, and F.~Posa, Eds., vol.
  7477.\hskip 1em plus 0.5em minus 0.4em\relax SPIE, 2009, p. 74770I.

\bibitem{Halimi2010}
A.~Halimi, Y.~Altmann, N.~Dobigeon, and J.-Y. Tourneret, ``Nonlinear unmixing
  of hyperspectral images using a generalized bilinear model,'' \emph{IEEE
  Trans. Geosci. and Remote Sensing}, vol.~49, no.~11, pp. 4153--4162, Nov.
  2011.

\bibitem{Chen2012}
J.~Chen, C.~Richard, and P.~Honeine, ``Nonlinear unmixing of hyperspectral data
  based on a linear-mixture/nonlinear-fluctuation model,'' \emph{IEEE Trans.
  Signal Process.}, vol.~61, no.~2, pp. 480--492, 2013.

\bibitem{Altmann2014a}
Y.~Altmann, N.~Dobigeon, S.~McLaughlin, and J.-Y. Tourneret, ``Residual
  component analysis of hyperspectral images - application to joint nonlinear
  unmixing and nonlinearity detection,'' \emph{IEEE Trans. Image Processing},
  vol.~23, no.~5, pp. 2148--2158, May 2014.

\bibitem{Dobigeon_IEEE_WHISPERS_2013}
N.~Dobigeon and C.~F\'evotte, ``Robust nonnegative matrix factorization for
  nonlinear unmixing of hyperspectral images,'' in \emph{Proc. IEEE GRSS
  Workshop on Hyperspectral Image and SIgnal Processing: Evolution in Remote
  Sensing (WHISPERS)}, Gainesville, FL, June 2013.

\bibitem{Newstadt2014ssp}
G.~E. Newstadt, A.~H. {III}, and J.~Simmons, ``Robust spectral unmixing for
  anomaly detection,'' in \emph{Proc. IEEE-SP Workshop Stat. and Signal
  Processing}, Gold Coast, Australia, July 2014.

\bibitem{Zare2014}
A.~Zare and K.~Ho, ``Endmember variability in hyperspectral analysis:
  Addressing spectral variability during spectral unmixing,'' \emph{IEEE Signal
  Processing Magazine}, vol.~31, no.~1, pp. 95--104, Jan 2014.

\bibitem{Pereyra2014ssp}
M.~Pereyra, N.~Whiteley, C.~Andrieu, and J.-Y. Tourneret, ``Maximum marginal
  likelihood estimation of the granularity coefficient of a {P}otts-{M}arkov
  random field within an mcmc algorithm,'' in \emph{Proc. IEEE-SP Workshop
  Stat. and Signal Processing}, Gold Coast, Australia, July 2014.

\bibitem{Nascimento2012}
J.~Nascimento and J.~Bioucas-Dias, ``Hyperspectral unmixing based on mixtures
  of dirichlet components,'' \emph{IEEE Trans. Geosci. and Remote Sensing},
  vol.~50, no.~3, pp. 863--878, March 2012.

\bibitem{Eches2011}
O.~Eches, N.~Dobigeon, and J.-Y. Tourneret, ``Enhancing hyperspectral image
  unmixing with spatial correlations,'' \emph{IEEE Trans. Geosci. and Remote
  Sensing}, vol.~49, no.~11, pp. 4239--4247, Nov. 2011.

\bibitem{Du2014}
X.~Du, A.~Zare, P.~Gader, and D.~Dranishnikov, ``Spatial and spectral unmixing
  using the beta compositional model,'' \emph{IEEE J. Sel. Topics Appl. Earth
  Observations Remote Sensing}, vol.~7, no.~6, pp. 1994--2003, June 2014.

\bibitem{Dobigeon2009}
N.~Dobigeon, S.~Moussaoui, M.~Coulon, J.-. Tourneret, and A.~O. Hero, ``Joint
  {B}ayesian endmember extraction and linear unmixing for hyperspectral
  imagery,'' \emph{IEEE Trans. Signal Process.}, vol.~57, no.~11, pp.
  2657--2669, Nov. 2009.

\bibitem{Altmann2014b}
Y.~Altmann, N.~Dobigeon, and J.-Y. Tourneret, ``Unsupervised post-nonlinear
  unmixing of hyperspectral images using a {H}amiltonian {M}onte {C}arlo
  algorithm,'' \emph{IEEE Trans. Image Processing}, vol.~23, no.~6, pp.
  2663--2675, June 2014.

\bibitem{Pereyra2013ip}
M.~Pereyra, N.~Dobigeon, H.~Batatia, and J.-Y. Tourneret, ``Estimating the
  granularity coefficient of a {P}otts-{M}arkov random field within an {MCMC}
  algorithm,'' \emph{IEEE Trans. Image Processing}, vol.~22, no.~6, pp.
  2385--2397, June 2013.

\bibitem{Robert2004}
C.~P. Robert and G.~Casella, \emph{Monte Carlo Statistical Methods},
  2nd~ed.\hskip 1em plus 0.5em minus 0.4em\relax New York: Springer-Verlag,
  2004.

\bibitem{Pakman2012}
A.~{Pakman} and L.~{Paninski}, ``{Exact Hamiltonian Monte Carlo for Truncated
  Multivariate Gaussians},'' \emph{ArXiv e-prints}, Aug. 2012.

\bibitem{Ding2011}
X.~Ding, L.~He, and L.~Carin, ``Bayesian robust principal component analysis,''
  \emph{IEEE Trans. Image Processing}, vol.~20, no.~12, pp. 3419--3430, Dec.
  2011.

\bibitem{Nascimento2005}
J.~M. Nascimento and J.~M. {Bioucas-Dias}, ``Vertex component analysis: A fast
  algorithm to unmix hyperspectral data,'' \emph{IEEE Trans. Geosci. and Remote
  Sensing}, vol.~43, no.~4, pp. 898--910, April 2005.

\bibitem{Fan2009}
W.~Fan, B.~Hu, J.~Miller, and M.~Li, ``Comparative study between a new
  nonlinear model and common linear model for analysing laboratory
  simulated-forest hyperspectral data,'' \emph{Remote Sensing of Environment},
  vol.~30, no.~11, pp. 2951--2962, June 2009.

\bibitem{Winter1999}
M.~Winter, ``Fast autonomous spectral end-member determination in hyperspectral
  data,'' in \emph{Proc. 13th Int. Conf. on Applied Geologic Remote Sensing},
  vol.~2, Vancouver, April 1999, pp. 337--344.

\bibitem{Altmann_RBLU_arxiv2015}
Y.~Altmann, S.~McLaughlin, and A.~Hero, ``{Robust Linear Spectral Unmixing
  using Anomaly Detection},'' \emph{ArXiv e-prints}, May 2015, available at
  http://arxiv.org/abs/1501.03731.

\bibitem{Besson_IEEE_Trans_SP_2011}
O.~Besson, N.~Dobigeon, and J.-Y. Tourneret, ``Minimum mean square distance
  estimation of a subspace,'' \emph{IEEE Trans. Signal Processing}, vol.~59,
  no.~12, pp. 5709--5720, Dec. 2011.

\bibitem{Eches2010a}
O.~Eches, N.~Dobigeon, C.~Mailhes, and J.-Y. Tourneret, ``Bayesian estimation
  of linear mixtures using the normal compositional model,'' \emph{IEEE Trans.
  Image Processing}, vol.~19, no.~6, pp. 1403--1413, June 2010.

\bibitem{Eches2010}
O.~Eches, N.~Dobigeon, and J.-Y. Tourneret, ``Estimating the number of
  endmembers in hyperspectral images using the normal compositional model an a
  hierarchical bayesian algorithm,'' \emph{IEEE J. Sel. Topics Signal
  Processing}, vol.~3, no.~3, pp. 582--591, June 2010.

\bibitem{Sheeren2011}
D.~Sheeren, M.~Fauvel, S.~Ladet, A.~Jacquin, G.~Bertoni, and A.~Gibon,
  ``Mapping ash tree colonization in an agricultural mountain landscape:
  Investigating the potential of hyperspectral imagery,'' in \emph{Proc. IEEE
  Int. Conf. Geosci. and Remote Sensing (IGARSS)}, July 2011, pp. 3672--3675.

\bibitem{Altmann2013}
Y.~Altmann, N.~Dobigeon, S.~McLaughlin, and J.~Tourneret, ``Nonlinear spectral
  unmixing of hyperspectral images using {G}aussian processes,'' \emph{IEEE
  Trans. Signal Process.}, vol.~61, no.~10, pp. 2442--2453, May 2013.

\end{thebibliography}

\end{document}